# Is There Quantum Recurrence in the Presence of an Energy Continuum?

James P. Lavine
Department of Physics
Georgetown University
37th and O Sts. NW
Washington, DC 20057 USA

## Abstract

Suppose an initial state is coupled to a continuum of energy states. The population of the initial state is expected to decrease with time, but is the decrease monotonic? The occupation probability of the initial state is the survival probability and the question is equivalent to asking if there are intervals of time where the survival probability increases. Such regrowth is also referred to as regeneration or recurrence and it occurs in systems with a countable number of discrete states. Regrowth is investigated with a simple model that allows transitions between the initial state and continuum states, but transitions between continuum states are not permitted. The model uses the solution of Schrödinger's Equation for a full energy continuum. Such a continuum runs from -∞ to +∞ and is found to have only exponential decay in time. However, the survival probability for a truncated continuum turns out to have a wide variety of behaviors. Generally, the survival probability decreases by several orders of magnitude, often as an exponential, and then has limited regrowth.

## 1. Introduction

If we consider the decay of an initially occupied state *s* into a finite set of states, then the Quantum Recurrence Theorem, QRT, applies [1-3]. The QRT says that if one picks a time *t* with the survival probability of state s, $p_s(t)$, then there exists another time $T$ such that $|p_s(t) - p_s(T)|$ is arbitrarily small. Such recurrences rely on interference between the allowed transitions of the system. Recently, I became curious about regeneration or regrowth when the energy spectrum has a continuum. I consulted references [1-3] and all three assume a discrete set of states. Bocchieri and Loinger [1] make a comment in passing that the QRT does not apply to systems with a continuum of energies. Perhaps this may be viewed in the following informal manner. The initial state decays to one of a large number of states that are very close in energy. Successive instances of decay go to a different one of these states, so no accumulation of probability occurs for a specific state. This is contrary to what happens with discrete states. It is as if the decay to these continuum states resembles a process with no memory. Hence, the decay of the survival probability is expected to be an exponential in time [4]. And, indeed, this is demonstrated below when the energy continuum runs from $-\infty$ to $+\infty$. However, the exponential is not found to be guaranteed for a truncated continuum.

The following sections provide derivations and numerical calculations that bear on the nature of regrowth when a continuum of energy states is present. The computations are based on a model introduced by Bixon and Jortner [5] to treat intramolecular radiationless transitions in an isolated molecule. This model has a discrete set of states and its closed-form solution leads to a sum over the states. This sum may be turned into an integral over energy and this integral is the basis of the numerical work presented here. In addition, this integral provides the full solution of Schrödinger's Equation when the energy ranges from $-\infty$ to $+\infty$, which is referred to as a full continuum. Facchi [6] works through this solution in his Section 3.4. Two more continua are treated. One introduces a lower bound on the energy and is a truncated continuum, while the other has a lower energy bound and an upper energy bound and is called a doubly-truncated continuum. These last two continua are treated by putting their energy limits on the integral used for the full continuum. I emphasize this is an approximate approach.

Section 2 introduces the needed formalism for studying the decay of an initial state through the survival amplitude and the survival probability. Their behaviors at large times and at short times are considered with the help of Appendix A for the former. Appendices B and C are devoted to the derivation of the closed-form solution of the Bixon-Jortner model. Sections 3 to 5 have the numerical results for the full continuum, the truncated continuum, and the doubly-truncated continuum, respectively. Appendix D shows the solution for the full continuum and Appendix E considers some of the oscillatory behavior seen in the survival probability versus time for the doubly-truncated continuum. The final section, Section 6, discusses what has been learned and what may be attacked next.

## 2. Formalism, Limits, and the Model

This section defines the survival amplitude, $A_s(t)$, and the survival probability, $p_s(t)$. Some general properties of $A_s(t)$ are developed since these bear on whether recurrence occurs. Next, the behavior of $A_s(t)$ and $p_s(t)$ with the time $t$ are considered. The large time limit is explored in Appendix A, while the $t \to 0$ limit is discussed in this section. The latter is related to the finiteness of the energy moments of the initial state. Finally, the Bixon-Jortner model is introduced with a continuum of energy states.

The system starts in state $|s\rangle$ and this is also written as $|\psi_s^0\rangle$, with $\langle\psi_s^0|\psi_s^0\rangle = 1$. The survival amplitude is defined to be

$$A_s(t) = \langle\psi_s^0|\psi(t)\rangle = \langle\psi_s^0|\hat{U}(t,0)|\psi_s^0\rangle, \tag{1}$$

where $\hat{U}(t,0)$ is the time-evolution operator. The present Hamiltonian $\hat{H}$ is assumed to be independent of time and $\hat{H} = \hat{H_0} + \hat{V}$ with $\hat{V}$ causing the transitions within the system. Hence [7],

$$A_s(t) = \langle\psi_s^0|\hat{U}(t,0)|\psi_s^0\rangle = \langle\psi_s^0|exp(-i\hat{H}t/\hbar)|\psi_s^0\rangle. \tag{2}$$

The survival probability is

$$p_s(t) = A_s(t)^*A_s(t). \tag{3}$$

Next, I define

$$exp(-i\hat{H}t/\hbar)|\psi_s^0\rangle = A_s(t)|\psi_s^0\rangle + |\varphi_d(t)\rangle. \tag{4}$$

The first term on the right-hand side represents the survival of the initial state and the second term accounts for the decays from the initial state. The key point about Eq. (4) is

$$\langle\psi_s^0|\varphi_d(t)\rangle = 0, \tag{5}$$

and this follows when Eq. (4) is multiplied from the left by $\langle\psi_s^0|$ , and Eq. (2) is used with $\langle\psi_s^0|\psi_s^0\rangle = 1$ . We continue with the help of Fonda and Ghirardi [8] and Fleming [9] to explore $A_s(t)$.

We start with

$$exp(-i\widehat{H}(t+t')/\hbar)|\psi_s^0\rangle = A_s(t+t')|\psi_s^0\rangle + |\varphi_d(t+t')\rangle , \qquad (6)$$

and we may multiply Eq. (4) by $exp(-i\widehat{H}t'/\hbar)$ to find

$$exp(-i\widehat{H}(t+t')/\hbar)|\psi_s^0\rangle = exp(-i\widehat{H}t'/\hbar)\{A_s(t)|\psi_s^0\rangle + |\varphi_d(t)\rangle\} =$$

$$A_s(t)exp(-i\widehat{H}t'/\hbar)|\psi_s^0\rangle + exp(-i\widehat{H}t'/\hbar)|\varphi_d(t)\rangle =$$

$$A_s(t)A_s(t')|\psi_s^0\rangle + A_s(t)|\varphi_d(t')\rangle + exp(-i\widehat{H}t'/\hbar)|\varphi_d(t)\rangle . \qquad (7)$$

Here Eq. (4) has been used in going from the first line to the third line. This works because the exponential operator does not affect $A_s(t)$. The right-hand sides of Eqs. (6) and (7) are equal, so when we multiply from the left with $\langle\psi_s^0|$, we are left with

$$A_s(t+t') = A_s(t)A_s(t') + \langle\psi_s^0| exp(-i\widehat{H}(t')/\hbar)|\varphi_d(t)\rangle . \qquad (8)$$

And this is our key result! If we drop the second term on the right-hand side, which is sometimes denoted $R_s(t',t)$, then we have

$$A_s(t+t') = A_s(t)A_s(t') . \qquad (9)$$

This is satisfied when the survival amplitude is an exponential in time multiplied by a possible phase factor. Then the survival probability is also an exponential in time, since the possible phase factors disappear when Eq. (3) is used. The flip side of this is that a nonzero second term in Eq. (8) means the survival probability is <u>not</u> an exponential in time! And, in addition, the presence of $R_s(t',t)$ means that regeneration occurs [4,10,11]. This is verified by my recent work [12] with the Bixon-Jortner model [5]. I used at most 26 energy levels and I found regeneration. In addition, some cases do show an exponential decay with time for $p_s(t)$ over a limited range starting at $t = 0$. So, for a finite number of energy levels, all is well. In fact,

Alexander [11] finds that an infinite number of discrete energy levels also leads to regeneration.

Hence, the question is what happens with a continuum of energy levels? The behavior at large times of $A_s(t)$ is covered in Appendix A, where the Riemann-Lebesgue Lemma and a theorem due to Paley and Wiener are discussed. These constrain $A_s(t)$ and rule out an exponential at large times for the truncated continuum.

What does the occupation probability $p_s(t)$ look like when the time $t$ approaches 0? References 3, 4, and 8 provide details. In brief, we have

$$A_s(t) = \langle \psi_s^0 | \widehat{U}(t,0) | \psi_s^0 \rangle = \langle \psi_s^0 | exp(-i\widehat{H}t/\hbar) | \psi_s^0 \rangle , \tag{10}$$

and

$$p_s(t) = A_s(t)^* A_s(t) . \tag{11}$$

When $t \to 0$, we expand the time evolution operator into a series in the time $t$ through terms quadratic in time. This leads to

$$A_s(t) = \left\langle \psi_s^0 \left| 1 + \left(-\frac{i\widehat{H}t}{\hbar}\right) + \frac{1}{2}\left(-\frac{i\widehat{H}t}{\hbar}\right)^2 \right| \psi_s^0 \right\rangle , \tag{12}$$

and

$$A_s^*(t) = \left\langle \psi_s^0 \left| 1 + \left(+\frac{i\widehat{H}t}{\hbar}\right) + \frac{1}{2}\left(+\frac{i\widehat{H}t}{\hbar}\right)^2 \right| \psi_s^0 \right\rangle . \tag{13}$$

We take the product of Eqs. (12) and (13), collect terms, and find to the lowest order in $t$ that

$$p_s(t) \approx 1 - \left( \langle \psi_s^0 | \widehat{H}^2 | \psi_s^0 \rangle - \langle \psi_s^0 | \widehat{H} | \psi_s^0 \rangle^2 \right) \frac{t^2}{\hbar^2} , \tag{14}$$

so, $p_s(t)$ is quadratic in the time for $t \to 0$, if the energy moments are finite. If $p_s(t)$ is an exponential in time, then

$$p_s(t) = exp(-\alpha t) , \tag{15}$$

with $\alpha$ obtained from the slope of $p_s(t)$. Equation (15) is linear in $t$ as $t \to 0$, for at small $t$

$$p_s(t) \approx 1 - \alpha t . \tag{16}$$

Again, if the moments of the energy are finite, then Eq. (14) applies and $p_s(t)$ is quadratic for very short times. While Eq. (14) only has the first two moments, if either moment is infinite, then Eq. (14) does not apply. Thus, as we explore the three cases below, we need to compute the moments of the energy for the initial state. And, we note the predicted slopes [9,13] of $p_s(t)$ at $t = 0$. Equation (14) leads to a slope of zero at $t = 0$, while Eq. (16) predicts a slope of $-\alpha$ at $t = 0$.

Finally, it is time to introduce the model that I use here. The simplest case is for a continuum from $-\infty$ to $\infty$, and the Bixon-Jortner model [5] provides a path to investigate $A_s(t)$ for this case. I start with Eq. (27) of Luu and Ma [14]. (I also provide a derivation in the present Appendix B.)

$$A_s(t) = \langle |\psi_s^0|\psi(t)\rangle = \tag{17}$$

$$exp(-iE_st/\hbar)\left(\frac{1}{2\pi}\right)\int_{-\infty}^{\infty} d\omega \left\{(2(\pi)V^2/(\varepsilon\hbar))exp(-i\omega t)\right\}/\left\{\omega^2 + \left(\frac{\pi V^2}{\varepsilon\hbar}\right)^2\right\}.$$

$E_s$ is the energy of the initial state, which is set to zero here, $V$ is the transition matrix element value and $\varepsilon$ is the separation of the discrete energy states in the discrete formulation of the Bixon-Jortner model. For the present purposes, I treat $\pi V^2/\varepsilon\hbar$ as a parameter. Transitions are only allowed between the initial state $s$ and any of the continuum states. Transitions between continuum states are not allowed. Finally, $E = \hbar\omega$ is the energy of a state. A term of $V^2$ has been dropped from the denominator of Eq. (17). Some of the calculations do include this factor and this is stated when it occurs.

I note that models with these transition matrix element rules occur frequently. For example, in Pietenpol's [15] solvable model with a full continuum.

The point is Eq. (17) is a Fourier transform of a Lorentzian that results in

$$A_s(t) = \langle|\psi_s^0|\psi(t)\rangle = exp(-iE_s t/\hbar)exp(-\pi V^2 t/\varepsilon\hbar) \ . \tag{18}$$

This exponential may also be directly derived from Schrodinger's Equation with the Hamiltonian that incorporates the above transition matrix element rules. This is found in [6] and in the present Appendix D. A return to the survival probability finds

$$p_s(t) = A_s(t)^* A_s(t) = exp(-2\pi V^2 t/\varepsilon\hbar) \ , \tag{19}$$

and an exponential in time results! I found this same exponential with only 26 energy levels and sometimes even fewer [12], but the exponential held only over a limited range of times. Equation (19) applies for all times when a full continuum is present and this means there is no regeneration nor recurrence due to Eq. (9) being satisfied. As a consequence,

$$\langle\psi_s^0|exp(-i\hat{H}t'/\hbar)|\varphi_d(t)\rangle = 0 \ . \tag{20}$$

Still, I pose the question: What is the relationship of the QRT to systems with a continuum of energies? And does the extent of the continuum matter? For example, what happens when the spectrum runs from $E_{min}$ to $+\infty$, which is the physically reasonable case.

Equation (17) provides a way to study this. I set $E_s = 0$, and pick

$$\alpha = \frac{2\pi V^2}{\varepsilon\hbar} = 0.2 \ . \tag{21}$$

When $\varepsilon = 0.1$, this corresponds to

$$\bar{V} = \frac{V}{\hbar} = 0.0564 \ , \tag{22}$$

And relates these parameters to the discrete energy level case. Three cases are considered. The first uses the integral in Eq. (17), which yields an exponential for $p_s(t)$ for all times. The second assumes a minimum energy, so the integral of Eq. (17) runs from $E_{min}$ to infinity, while the third has $E_{min}$ and $E_{max}$ and is a doubly-truncated continuum. Thus, the third case represents a continuum of finite extent. The same qualitative behavior for $p_s(t)$ is found whether or not $\bar{V}^2$ is present in the denominator of Eq. (17).

Hence, examples of both are provided. I emphasize that Eq. (17) provides only an approximate solution for the last two continua

## 3. The Continuum Goes from $-\infty$ to $+\infty$

As stated above, this case has $A_s(t)$ proportional to an exponential and with Eq. (11)

$$p_s(t) = A_s(t)^* A_s(t) = exp(-2\pi V^2 t/\varepsilon\hbar) = exp(-\alpha t) . \qquad (23)$$

We start by finding the moments of the Hamiltonian. Are they finite? Let the $n$-th moment be

$$\langle\psi_s^0|\hat{H}^n|\psi_s^0\rangle = \Sigma_j\langle\psi_s^0|\hat{H}^n|j\rangle\langle j|\psi_s^0\rangle = \Sigma_j\langle\psi_s^0|E_j^n|j\rangle\langle j|\psi_s^0\rangle =$$

$$\Sigma_j\langle\psi_s^0|j\rangle E_j^n|j\rangle\langle j|\psi_s^0\rangle = \Sigma_j a_j^* a_j E_j^n . \qquad (24)$$

Here
$$E_j = \hbar\omega_j' , \qquad (25)$$

are the eigenenergies of $\hat{H}$ and the eigenkets are the $|j\rangle$. In addition, we have used an analogy to Eq. (B9) of Appendix B along with Eq. (B6).

It is easier now to go to an integral over the angular frequency $\omega$ in place of $\omega_j'$. This results in

$$\frac{\langle\psi_s^0|\hat{H}^n|\psi_s^0\rangle}{\hbar^n} =$$

$$\left(\frac{1}{2\pi}\right)\int_{-\infty}^{\infty} d\omega \left\{(2(\pi)V^2/(\varepsilon\hbar))\omega^n\right\}/\left\{\omega^2 + \left(\frac{\pi V^2}{\varepsilon\hbar}\right)^2\right\}. \qquad (26)$$

We use the results from Appendix B, especially Eq. (B25), to obtain this form. The energy of state $s$ is set to zero and the $V^2$ term is dropped from the denominator, although its presence does not change the following argument. The key point is to take advantage of the odd and even property of the integral over $\omega$. For odd $n$, the energy moment is zero, while for even $n$, the energy moment is infinite. Hence, Eq. (14) does not apply and we check if Eq. (16) is accurate as $t \to 0$.

Equation (17) is evaluated to find the survival amplitude by using the Integrate command in Mathematica [16]. This serves as a check on the numerical method. Then the survival probability is found through Eq. (3). **Please note**: When $\bar{V} = 0.0$ is attached to a figure, it simply means the $V^2$ term is not added to the denominator of Eq. (17).

Figure 1 shows a semi-logarithmic plot of $p_s(t)$, the survival probability, for $\alpha = 0.2$ and $\bar{V} = 0.0$ with a time step of 0.1. $p_s(t)$ is exponential as expected and its slope gives 0.200 in agreement with Eqs. (15) and (21). In addition, the linear approximation of Eq. (16) is verified for times below $t = 0.02$ with a time step of 0.001, but is not shown here.

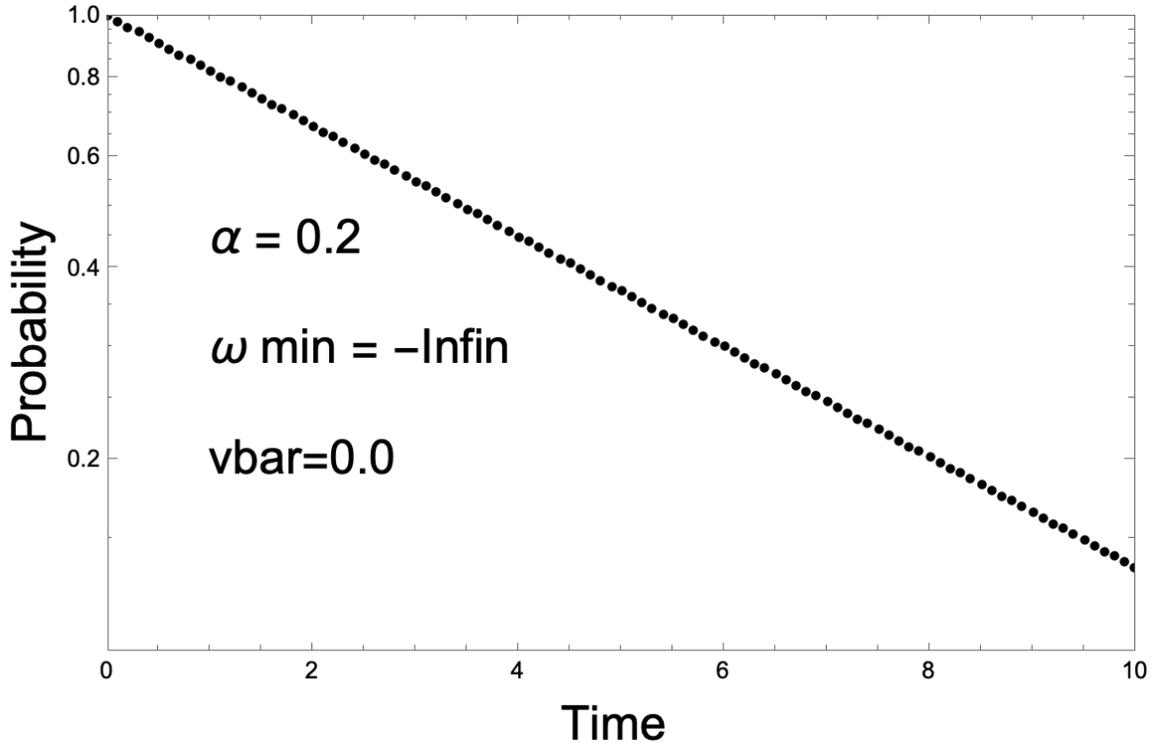

Figure 1. The survival probability versus time for the $-\infty$ to $+\infty$ continuum with $\alpha = 0.2$ and $\bar{V} = 0.0$. The vertical scale for this semi-logarithmic plot starts at 0.1.

We now turn to larger times for this case. Figure 2 has the real and imaginary parts of the survival amplitude. The imaginary part is zero due to the oddness of the integrand in Eq. (17) with the sine and the real part has not been normalized yet. The survival probability is presented in Fig. 3 in a semi-logarithmic plot. The exponential decay is clearly visible with a derived decay constant of 0.200, which is in agreement with the input value of $\alpha = 0.2$. Pietenpol [15] treats a decay model for an unstable state that incorporates the $-\infty$ to $+\infty$ continuum and has an energy-dependent transition matrix element. This case also yields a complete closed-form solution of Schrodinger's Equation. Pietenpol's model is a variant of the Lee model [17] and is discussed along with other full continuum models in Chapters 4 and 5 of [6].

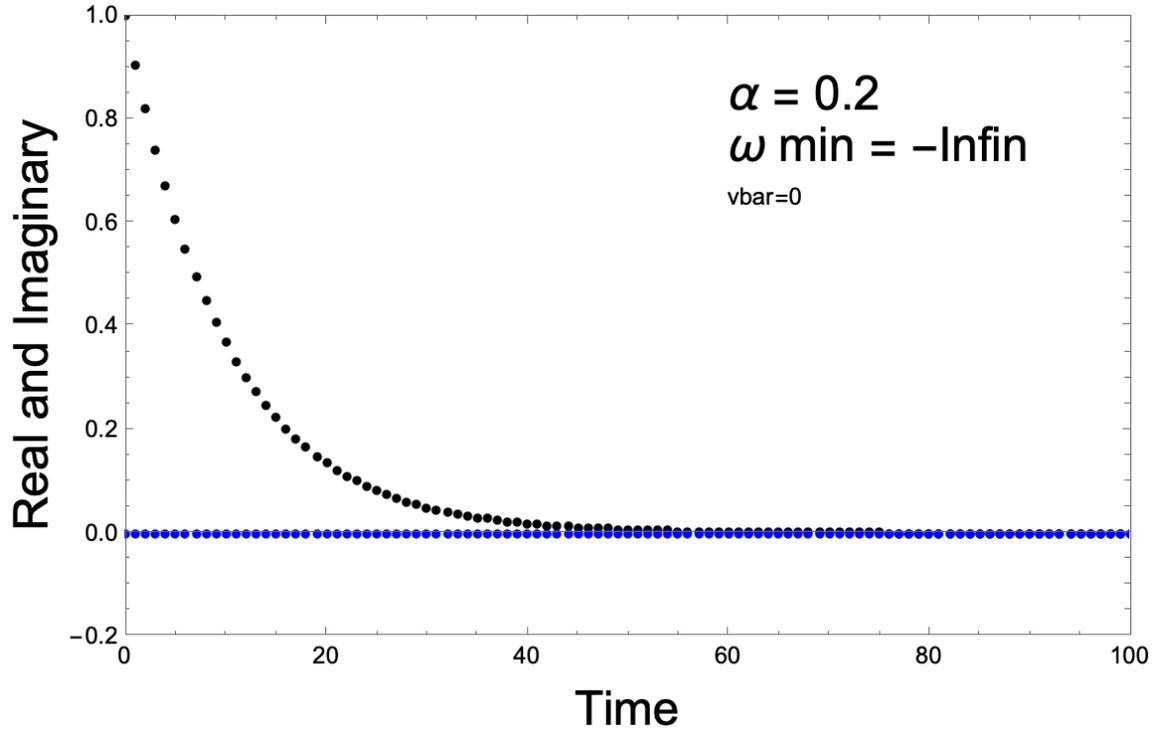

Figure 2. The survival amplitude's real (black) and imaginary (blue) parts versus time for the $-\infty$ to $+\infty$ continuum with $\alpha = 0.2$ and $\bar{V} = 0.0$.

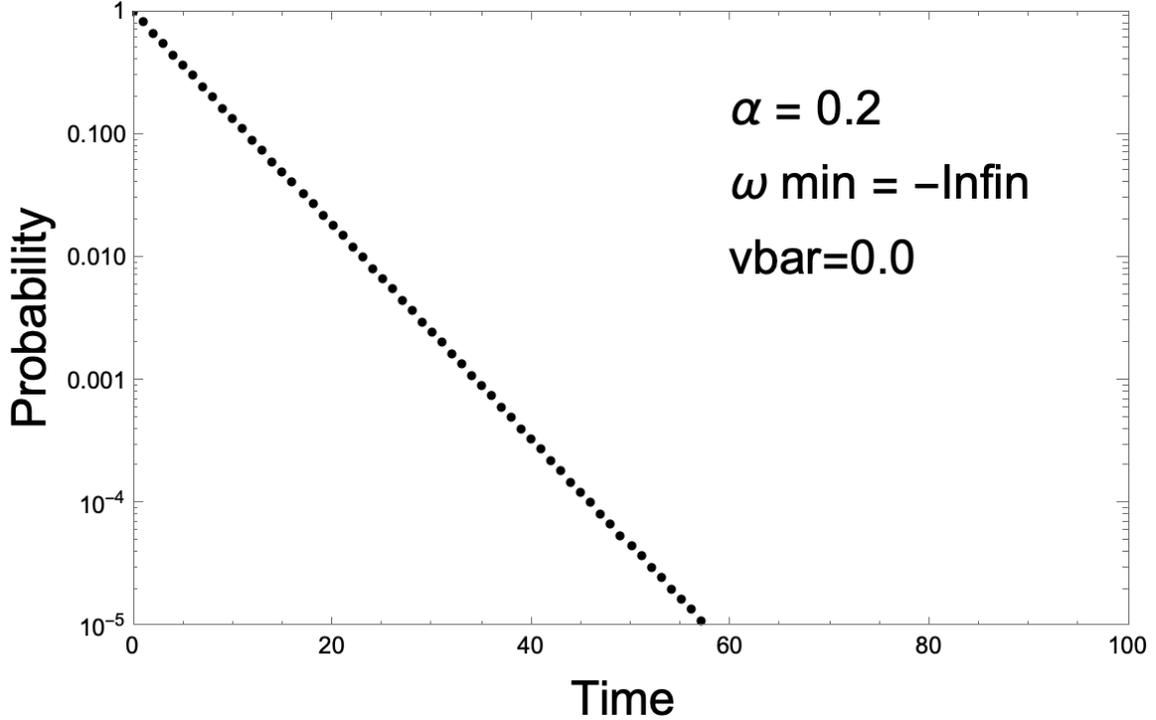

Figure 3. The survival probability versus time for the $-\infty$ to $+\infty$ continuum with $\alpha = 0.2$ and $\bar{V} = 0.0$.

## 4. The Continuum Goes from $\omega_{min}$ to $+\infty$

We next put a lower bound on the energy in order to treat a more realistic physical system, that is, one with a ground state. The integral in Eq. (17) now runs from $\omega_{min}$ to $+\infty$. The moments of the Hamiltonian are all infinite for this case, so we expect Eq. (16) to apply as $t \to 0$. Figure 4 is a linear plot of the survival probability $p_s(t)$ for $\omega_{min}=$ -0.2 and -1.0 with $\bar{V}$ present in the denominator of Eq. (17). The slope is constant for the first few data points, hence Eq. (16) applies, but then it starts to change. For these times, the $p_s(t)$ decreases slightly faster when $\omega_{min}$ goes to -0.2. However, this trend is reversed before the time reaches 1.

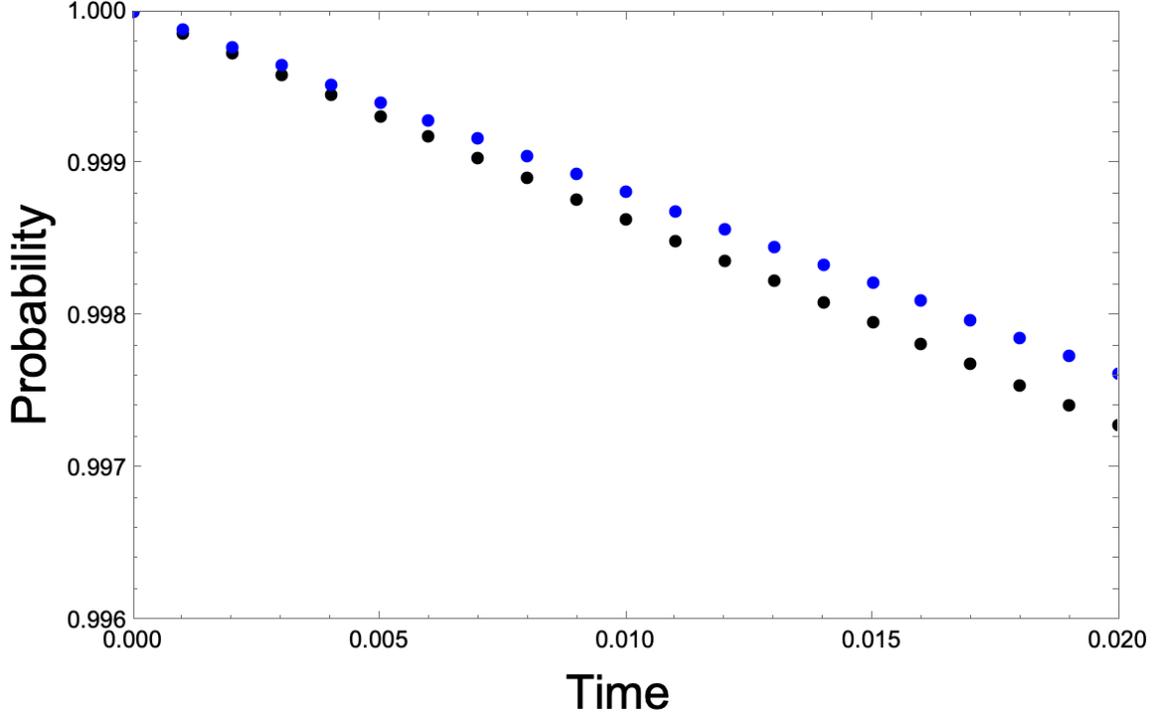

Figure 4. The survival probability versus time for the $\omega_{min}$ to $+\infty$ continuum with $\alpha = 0.2$, $\bar{V} = 0.0564$, and $\omega_{min}$=-1.0 (blue) and -0.2 (black). The vertical scale is linear and the time step is 0.001.

The larger time results start with Fig. 5 and the real and the imaginary parts of the unnormalized survival amplitude. As expected, with a finite $\omega_{min}$, the imaginary part in non-zero. These lead to the survival probability that is shown in Fig. 6. What is remarkable is the increase of $p_s(t)$ at $t > 20$ and again at $t > 50$. These rises may be termed regeneration or regrowth and are in accord with the consequences of the Paley-Wiener theorem discussed in Appendix A. It appears highly unlikely that recurrence in the sense of the quantum recurrence theorem will occur, that is, $p_s(t)$ returns to 1. In addition, the initial decay of $p_s(t)$ is not exponential in time in contrast to the case for the full continuum of Section 3. The same regrowth results are found with $\omega_{min} = $ -0.4 and -1.0, as seen in Figs 7 and 8, respectively.

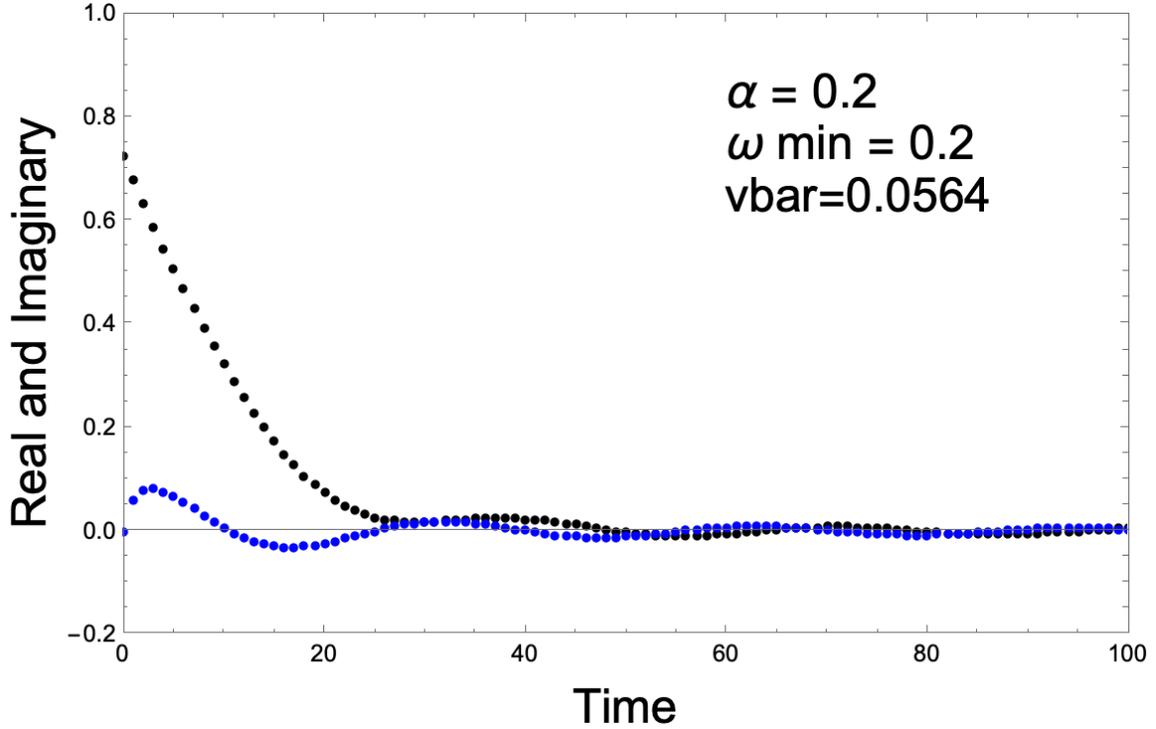

Figure 5. The real and imaginary parts of the unnormalized survival amplitude versus time for the $\omega_{min}$ to $+\infty$ continuum with $\alpha = 0.2, \bar{V} = 0.0564$, and $\omega_{min}$=-0.2. The label within the plot has the absolute value of $\omega_{min}$. The time step is 1.

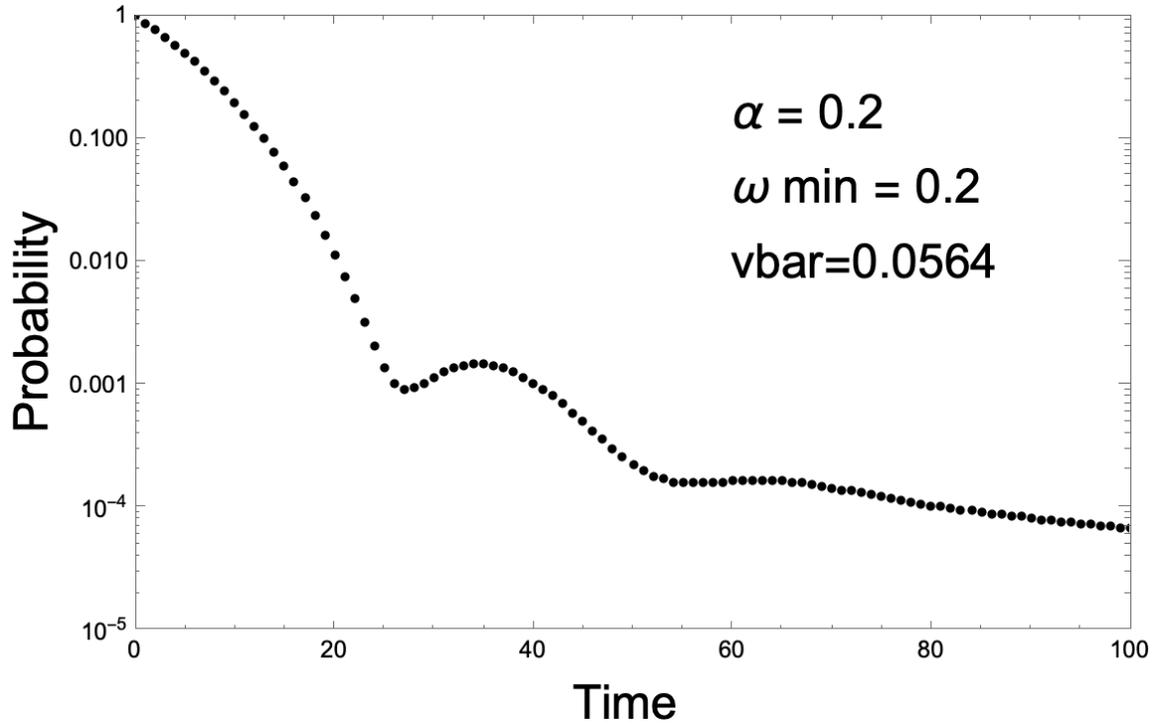

Figure 6. The survival probability versus time for the $\omega_{min}$ to $+\infty$ continuum with $\alpha = 0.2$, $\bar{V} = 0.0564$, and $\omega_{min}$=-0.2. The label within the plot has the absolute value of $\omega_{min}$. The time step is 1.

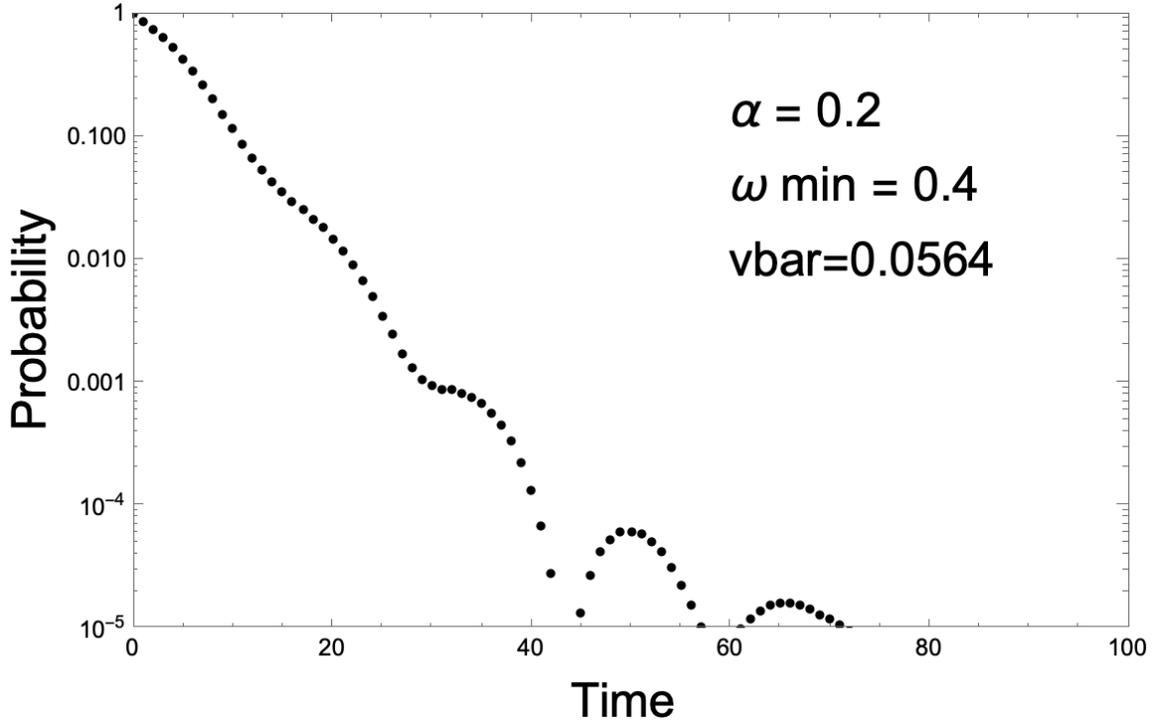

Figure 7. The survival probability versus time for the $\omega_{min}$ to $+\infty$ continuum with $\alpha = 0.2$, $\bar{V} = 0.0564$, and $\omega_{min}$=-0.4. The label within the plot has the absolute value of $\omega_{min}$. The time step is 1.

As the absolute value of $\omega_{min}$ increases, the fall of $p_s(t)$ becomes more exponential in appearance and closer to the behavior shown in Fig. 3 for the $-\infty$ to $+\infty$ continuum. The oscillations in Fig. 8 for $t > 40$ are investigated with a time step of 0.2. Three peaks appear as the time goes from 40 to 60, with $p_s(t)$ approximately equal to $10^{-6}$ at $t = 60$. The next time interval of 20 covers $t = 60$ to 80 and is shown in Fig. 9. The successive peaks decrease in size as the time grows larger and this continues through $t = 100$. These oscillations leave $p_s(80) \approx p_s(100)$. Then $p_s(t)$ hardly varies from about $1.2 \times 10^{-7}$ at t = 100 to just over $10^{-7}$ at $t = 120$.

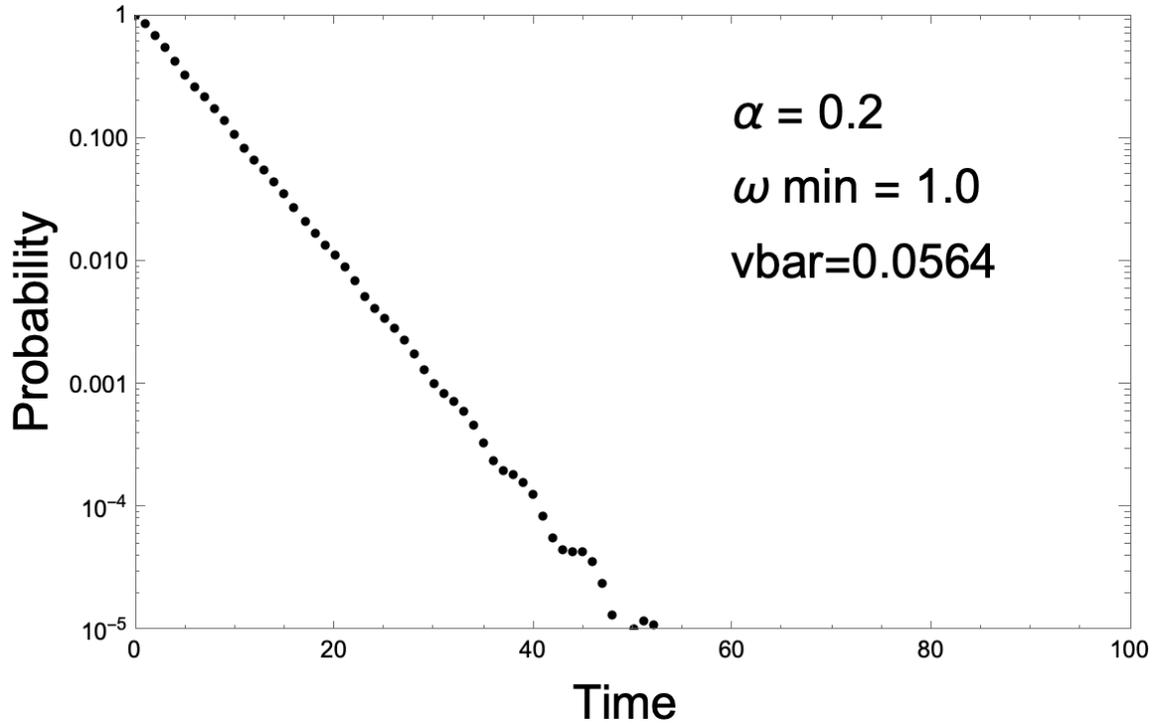

Figure 8. The survival probability versus time for the $\omega_{min}$ to $+\infty$ continuum with $\alpha = 0.2$, $\bar{V} = 0.0564$, and $\omega_{min}$=-1.0. The label within the plot has the absolute value of $\omega_{min}$. The time step is 1.

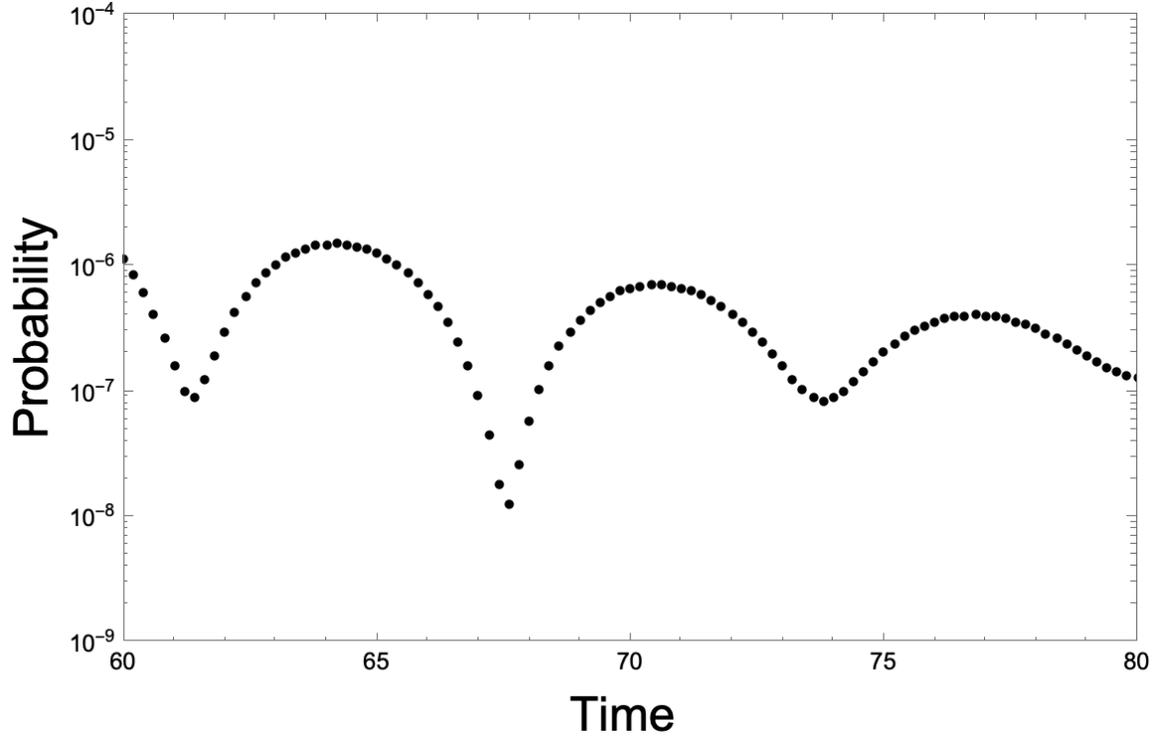

Figure 9. The survival probability versus time for the $\omega_{min}$ to $+\infty$ continuum with $\alpha = 0.2, \bar{V} = 0.0564$, and $\omega_{min}$=-1.0. The time step is 0.2 and the time goes from 60 to 80.

The calculations are extended to larger times and Fig. 10 presents the survival probability for $t$ = 140 to 180 with a time step of 0.4. This plot is a log-log plot and it shows that $p_s(t)$ is decreasing with an inverse power law, that is, the survival probability goes as $1/t^\delta$ with $\delta \sim 2.03$. We see the power law expected at large times for the case of the truncated continuum in the light of Eq. (A7) of Appendix A.

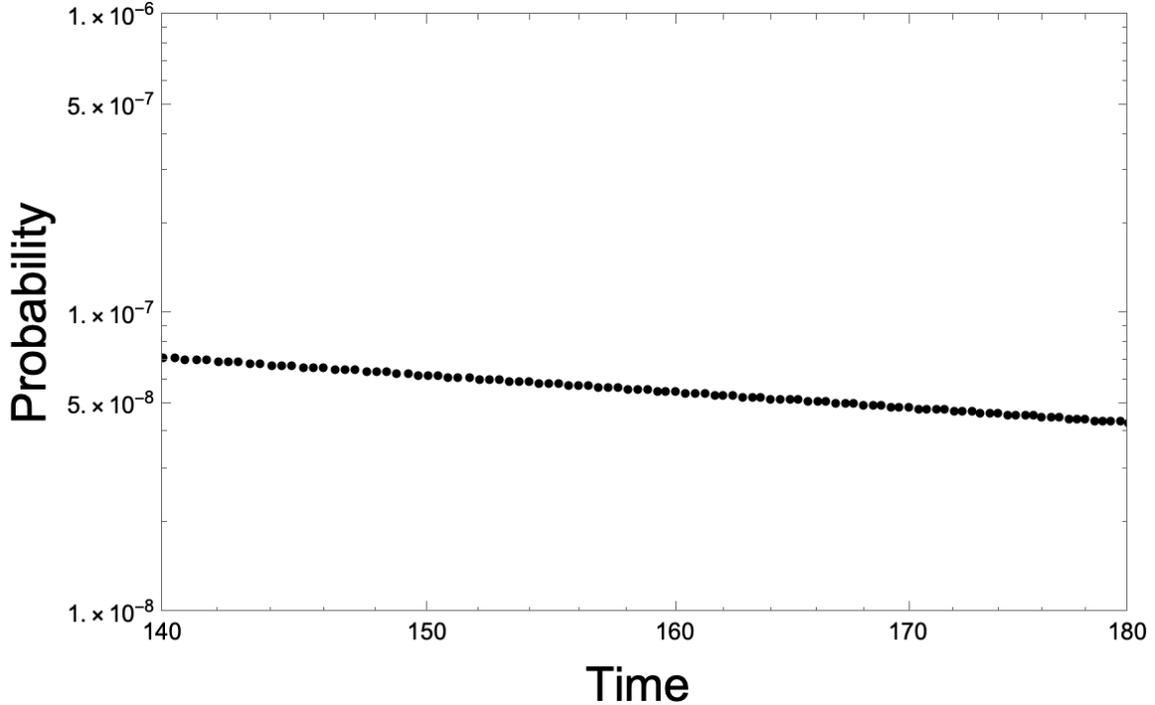

Figure 10. The survival probability versus time for the $\omega_{min}$ to $+\infty$ continuum with $\alpha = 0.2$, $\bar{V} = 0.0564$, and $\omega_{min} = -1.0$. The time step is 0.4 and the time goes from 140 to 180. This is a log-log plot.

As stated above, here the truncated continuum is coupled with a transition matrix element that only connects the initial state with states within the continuum. Continuum state to continuum state transitions are not considered. Interestingly, Cohen-Tannoudji, Diu, and Laloë [18] treat a similar situation in their Complement $D_{XIII}$ that covers the Wigner-Weisskopf approach to decay [19]. Nakazato, Namiki, and Pascazio delve into a related model for the survival amplitude in their section 3 [20]. They eventually find an exponential term and an inverse power law in time. Along the way they illustrate the complexities of performing inverse Laplace transforms through contour integrations. No calculations are included, but an example for this model occurs in Chapter 4 of the PhD thesis of Facchi [6].

Two investigations of a decaying system [21,22] with a truncated continuum show exponential decay followed by a series of oscillations that damp out to a slowly decreasing survival probability. The models have two contributions to the survival amplitude and their interference leads to the oscillations. This behavior is illustrated in Section 4 and Fig. 2 of Merlin [21], Fig. 3 of Onley and Kumar [22], and in Fig. 11, which shows the present author's exploration of the model of Ref. [21]. The magnitude of the second contribution is much less that the first. Hence, the second contribution is not easily seen until the first contribution has decreased significantly.

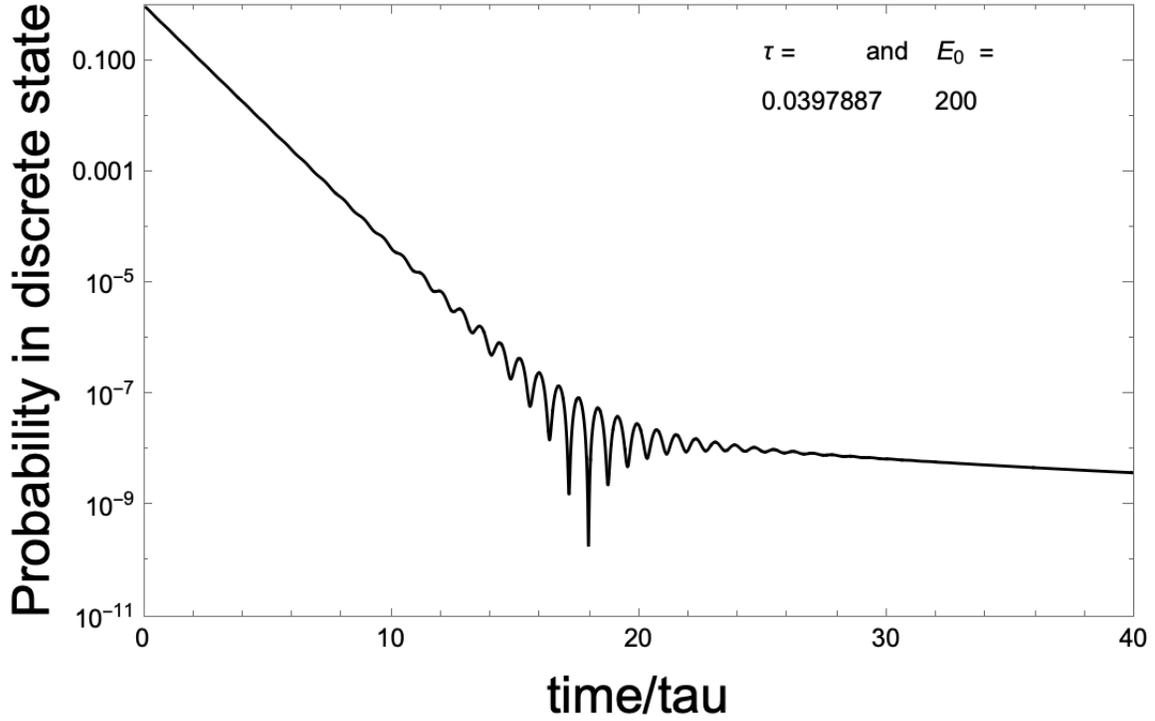

Figure 11. The survival probability calculated with the model of Merlin [21]. The exponential decay switches to an algebraic decay as the time increases. The oscillations occur when the two contributions to $p_s(t)$ are of similar magnitude and interfere.

The present work is based on the simpler Bixon-Jortner model and only has one mechanism, the transitions from the initial state to the continuum and back. While these interfere, each continuum state's population is very small and the effect is hard to display. However, this idea of interference is made visible for this truncated continuum model, when we rewrite Eq. (17) as two integrals. Here we represent part of the integrand by $M(\omega)$

$$\int_{-\beta}^{\infty} d\omega M(\omega) e^{-i\omega t} = \int_{-\infty}^{\infty} d\omega\, M(\omega) e^{-i\omega t} - \int_{-\infty}^{-\beta} d\omega M(\omega) e^{-i\omega t}. \qquad (27)$$

Now the first integral is real since the integral over the sine term is odd and vanishes. The second integral contributes both a real and an imaginary term.

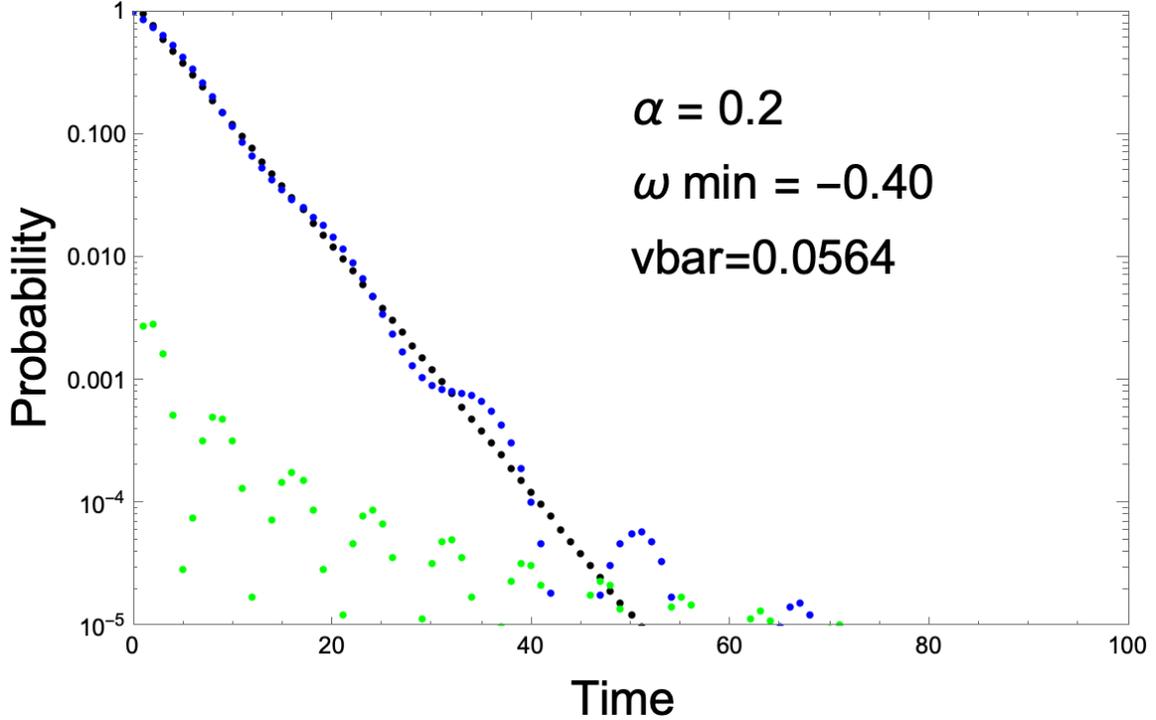

Figure 12. The components of the survival probability versus time for the $\omega_{min}$ to $+\infty$ continuum with $\alpha = 0.2$, $\bar{V} = 0.0564$, and $\omega_{min}$=-0.4. The time step is 1. The black dots form a straight line and are the exponential term based on the first integral in Eq. (27). The blue dots are the real contribution of the second integral and they show oscillations. The green dots are the oscillating imaginary contribution to $p_s(t)$.

The interference of the two real terms is demonstrated in Fig. 12 and is one way to reveal the source of the oscillations in $p_s(t)$ for this case with the truncated continuum. A closing and fanciful thought: The presence of the lower limit on the energy does perturb the continuum. Perhaps, the lower limit causes populations to "accumulate" in energy states near this lower limit. This would enhance the continuum to initial state transitions and give rise through interference to the oscillations seen in Figs. 6 to 10. These populations are small, so they are only seen when $p_s(t)$ is reduced by several orders of magnitude.

## 5. The Continuum Goes from $\omega_{min}$ to $\omega_{max}$

The third and last case features a continuum with a minimum and a maximum energy. Here, we first set $\omega_{max} = |\omega_{min}|$. The moments of the

Hamiltonian are all finite for this case, so we expect Eq. (14) to apply as $t \to 0$. Figure 13 shows the survival probability when Eq. (17) runs from -3. to +3 and the time step is 0.1. The slope clearly changes for $t < 2$ and this is confirmed with the linear plot displayed in Fig. 14. The curve bends as $t \to 0$ and the slope at $t = 0$ is zero in accord with Eq. (14).

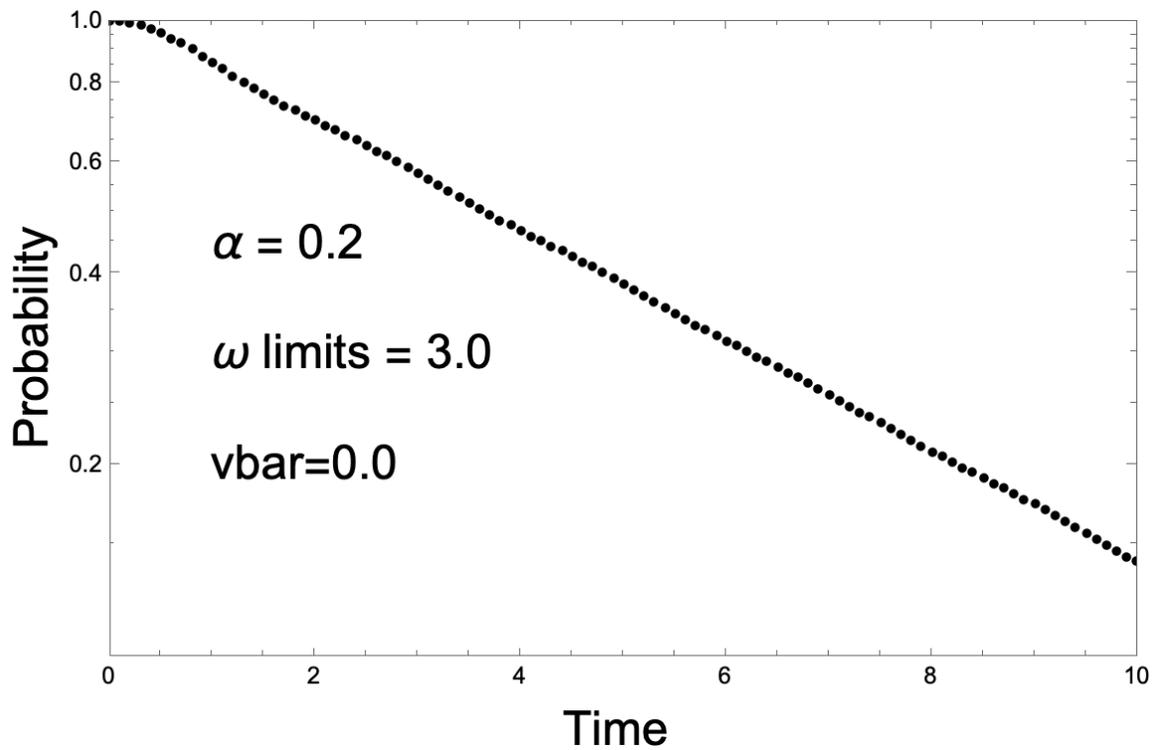

Figure 13. The survival probability versus time for the $\omega_{max} = 3 = |\omega_{min}|$ continuum with $\alpha = 0.2, \bar{V} = 0.0$. The time step is 0.1 and the vertical axis starts at 0.1 on this semi-logarithmic plot.

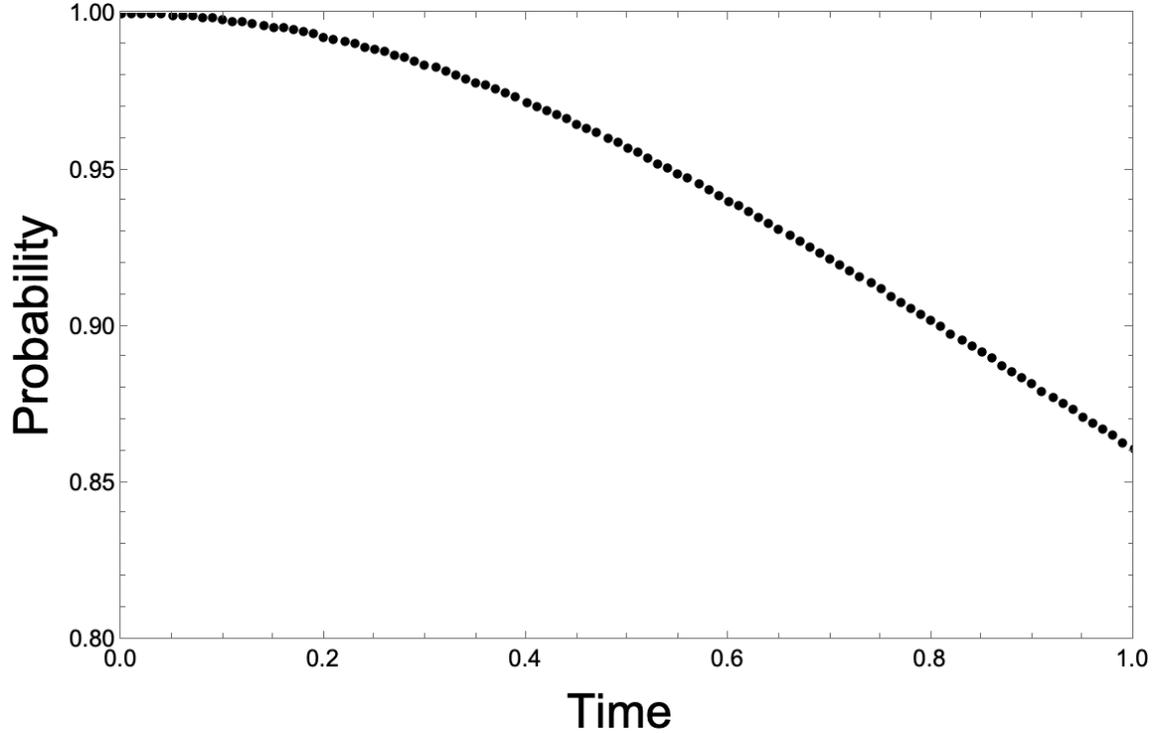

Figure 14. The survival probability versus time for the $\omega_{max} = 3 = |\omega_{min}|$ continuum with $\alpha = 0.2, \bar{V} = 0.0$. The time step is 0.01 and this is a linear plot.

We continue with Fig. 15 for larger times for the same limits of integration. This result is quite close to the exponential for the full continuum shown in Fig. 3 as might be expected for large $\omega$ limits. The plot yields 0.200 for the parameter for the exponential. The exponential decay trend of $p_s(t)$ continues below $10^{-5}$ as shown in Fig. 16. A rough exponential parameter is 0.19, while 0.18 is found if the peaks are used. Figure 16 also demonstrates that the amplitude of the oscillations grows when the time increases and additional calculations find they reach about two decades before $t \sim 95$.

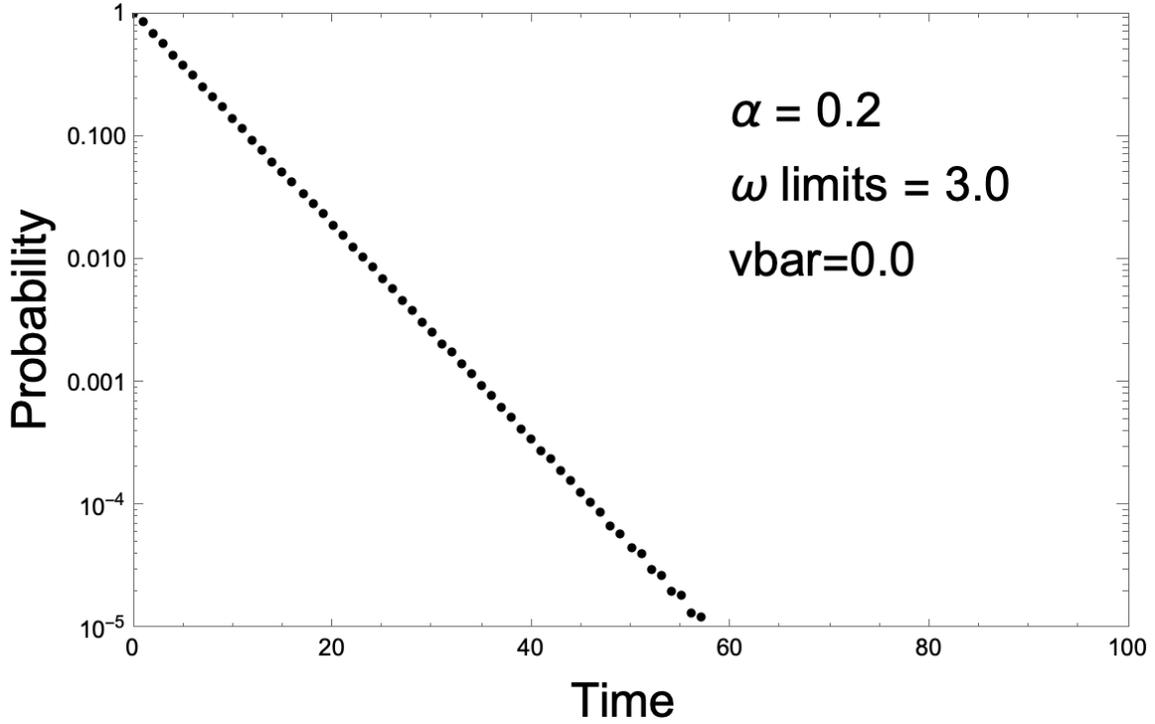

Figure 15. The survival probability versus time for the $\omega_{max} = 3 = |\omega_{min}|$ continuum with $\alpha = 0.2, \bar{V} = 0.0$. The time step is 1.0.

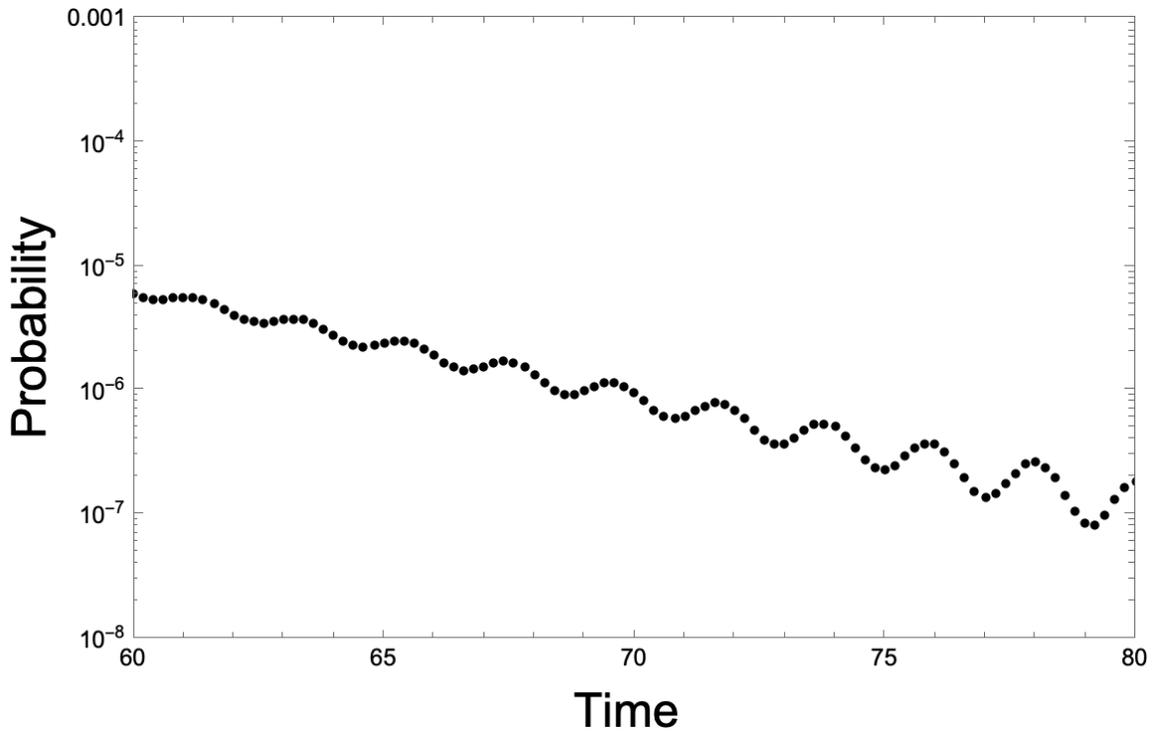

Figure 16. The survival probability versus time for the $\omega_{max} = 3 = |\omega_{min}|$ continuum with $\alpha = 0.2, \bar{V} = 0.0$. The time step is 0.2.

We next set $\omega_{max} = 1 = |\omega_{min}|$ and Fig. 17 shows strong deviations from an exponential. In addition, the survival probability again oscillates as the time increases with the swings increasing for times above 40. These variations in $p_s(t)$ are investigated in Figs. 18 and 19 for t from 55 to 75 and 75 to 95, respectively. Contrary to Fig, 16, larger times lead to continued oscillations with only a very gradual decrease in the peak amplitude as shown in Fig. 20.

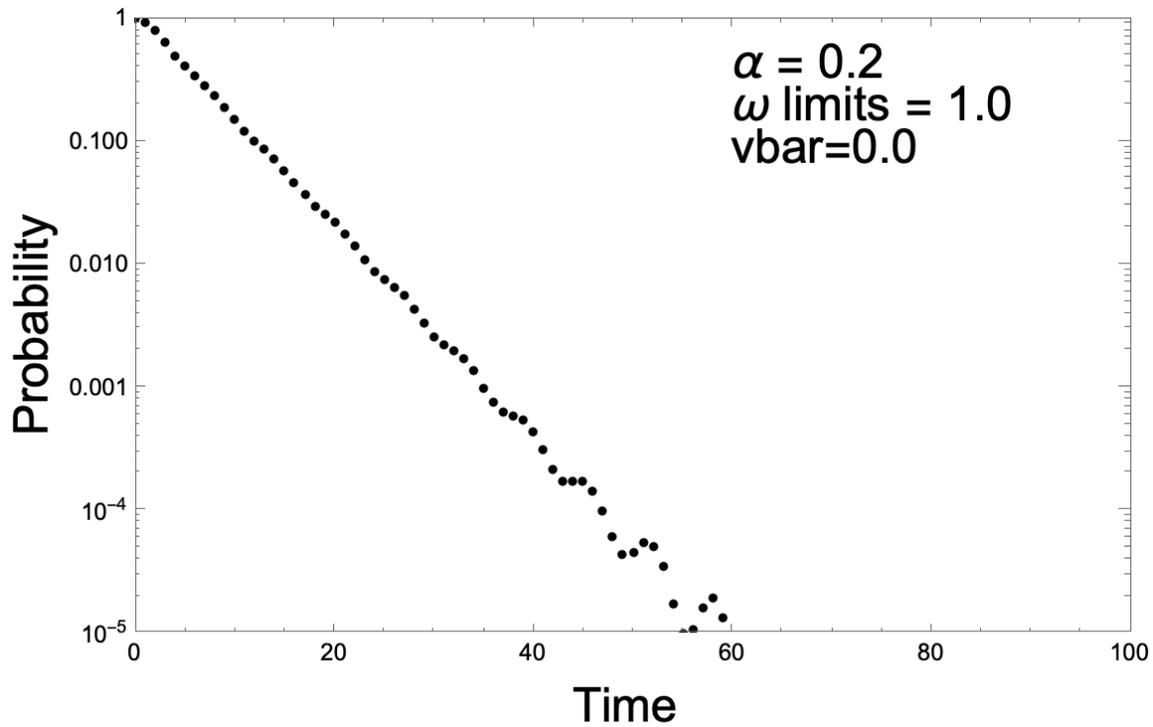

Figure 17. The survival probability versus time for the $\omega_{max} = 1 = |\omega_{min}|$ continuum with $\alpha = 0.2, \bar{V} = 0.0$. The time step is 1.0 and this is a semi-logarithmic plot.

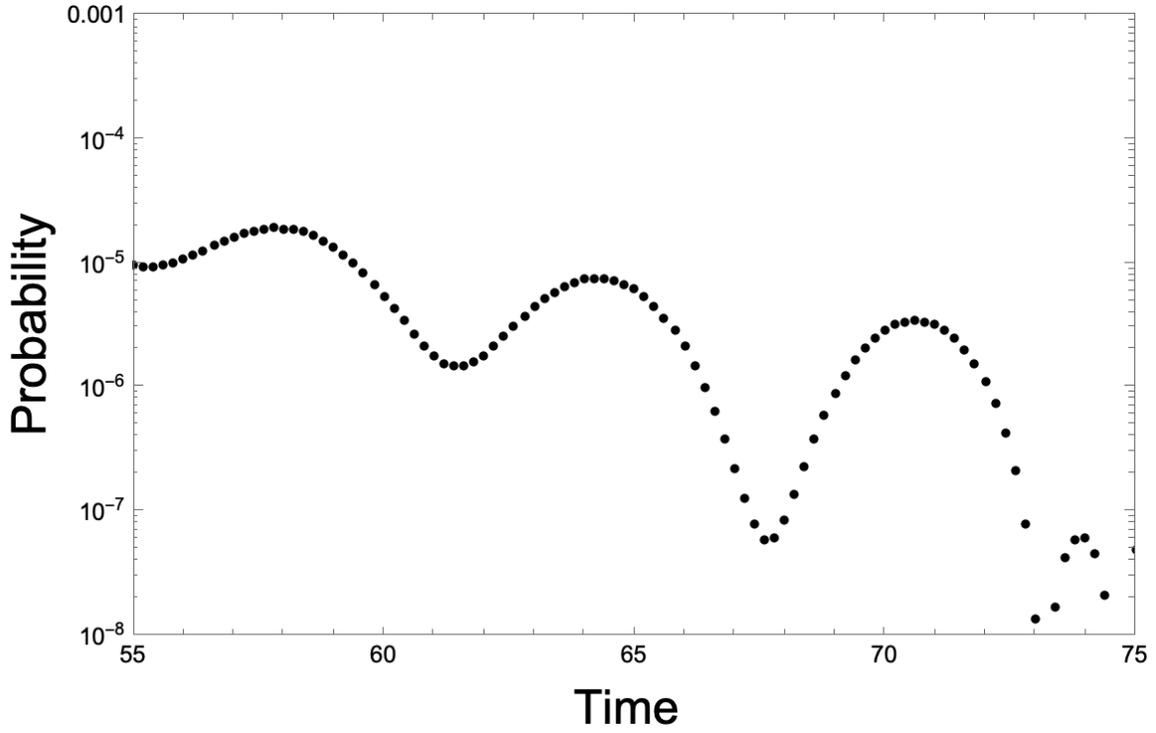

Figure 18. The survival probability versus time for the $\omega_{max} = 1 = |\omega_{min}|$ continuum with $\alpha = 0.2, \bar{V} = 0.0$. The time step is 0.2 and note the vertical axis.

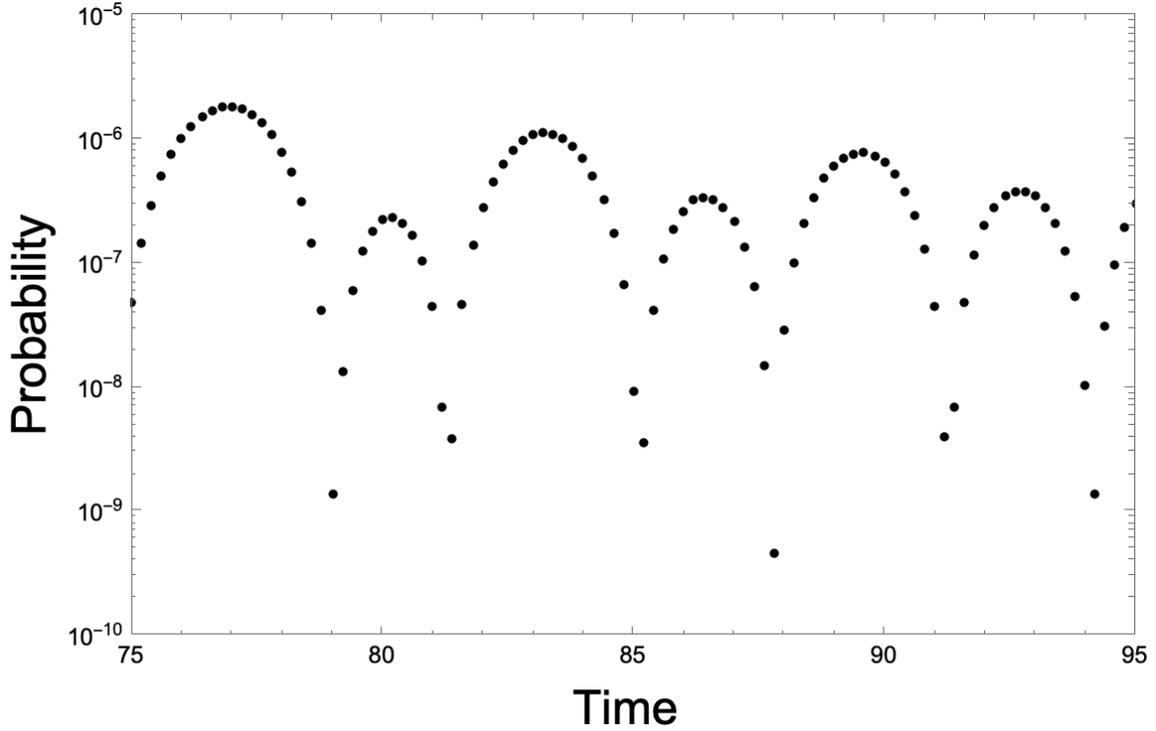

Figure 19. The survival probability versus time for the $\omega_{max} = 1 = |\omega_{min}|$ continuum with $\alpha = 0.2, \bar{V} = 0.0$. The time step is 0.2 and note the vertical axis.

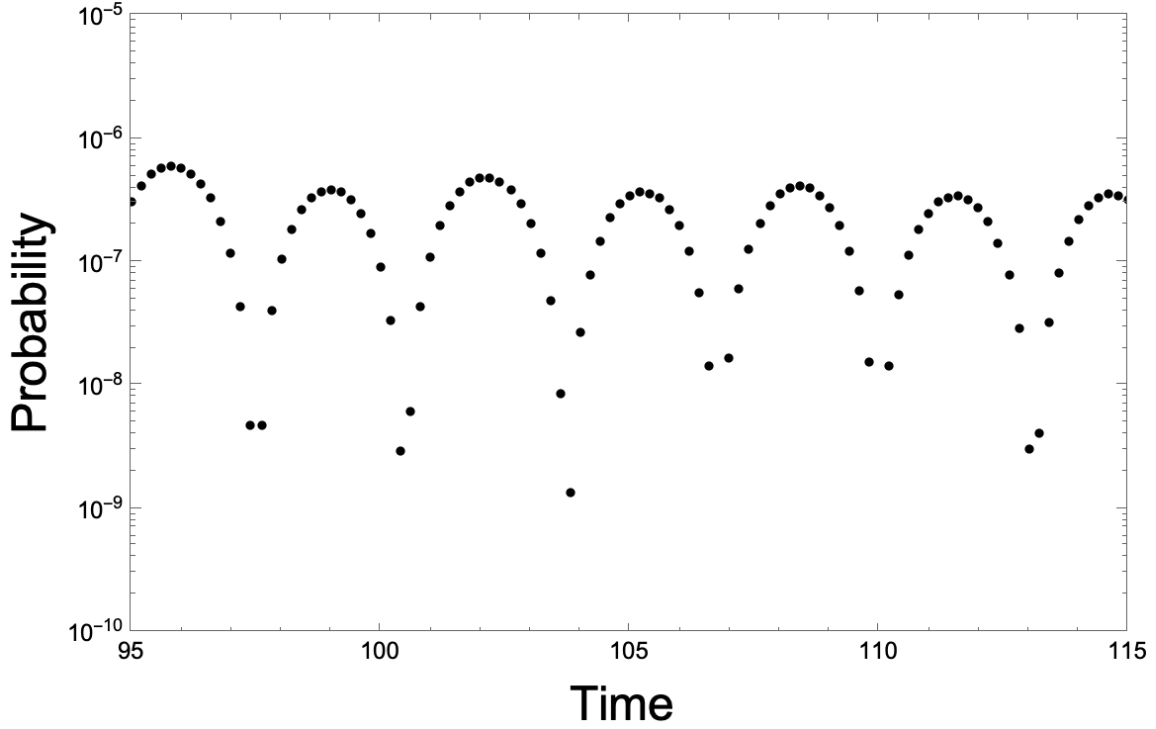

Figure 20. The survival probability versus time for the $\omega_{max} = 1 = |\omega_{min}|$ continuum with $\alpha = 0.2, \bar{V} = 0.0$. The time step is 0.2 and note the vertical axis.

Why these oscillations in the survival probability? I looked into the version of Eq. (17) for this case of the doubly-truncated continuum. The denominator of the integrand is broken into two pieces by the use of partial fractions and with $\delta = \alpha/2$ we find

$$\frac{1}{(\omega^2+\delta^2)} = (i/2\alpha)\{(1/(\omega + i\delta)) - (1/(\omega - i\delta))\}. \tag{28}$$

Since we have been treating the case with $\omega_{max} = |\omega_{min}|$, the integral over the sine in Eq. (17) yields zero. We are left with the cosine integral involving the two terms of Eq. (28). Each gives rise to a term with a cosine integral and a sine integral function leavened by a sinh and a cosh. The results are recorded in Appendix E. The two integral functions lead to the oscillations. Of course, in parallel with the truncated continuum, the two boundaries here may each be viewed as leading to an accumulation of population. Figure 20 then may be interpreted as the population exchange between an effective two-level system [23].

The natural question is what happens when $\omega_{max} \ne |\omega_{min}|$? I checked two cases with $\omega_{min} = -1$ and $\omega_{max} = 2$ or $3$. For each case, the initial decay resembles those in Figs. 15 and 17, but very slight undulations are present unlike the former. The decay from t = 60 to 80 differs from that in Fig. 16 as shown in Fig. 21, but does bear some similarity to Fig. 18. However, as the time gets larger, the oscillations become more frequent with smaller amplitudes. Figure 22 presents the survival probability $p_s(t)$ for $t$ = 180 to 200. There is now more resemblance to Fig. 16, although the peak amplitudes hardly decrease now.

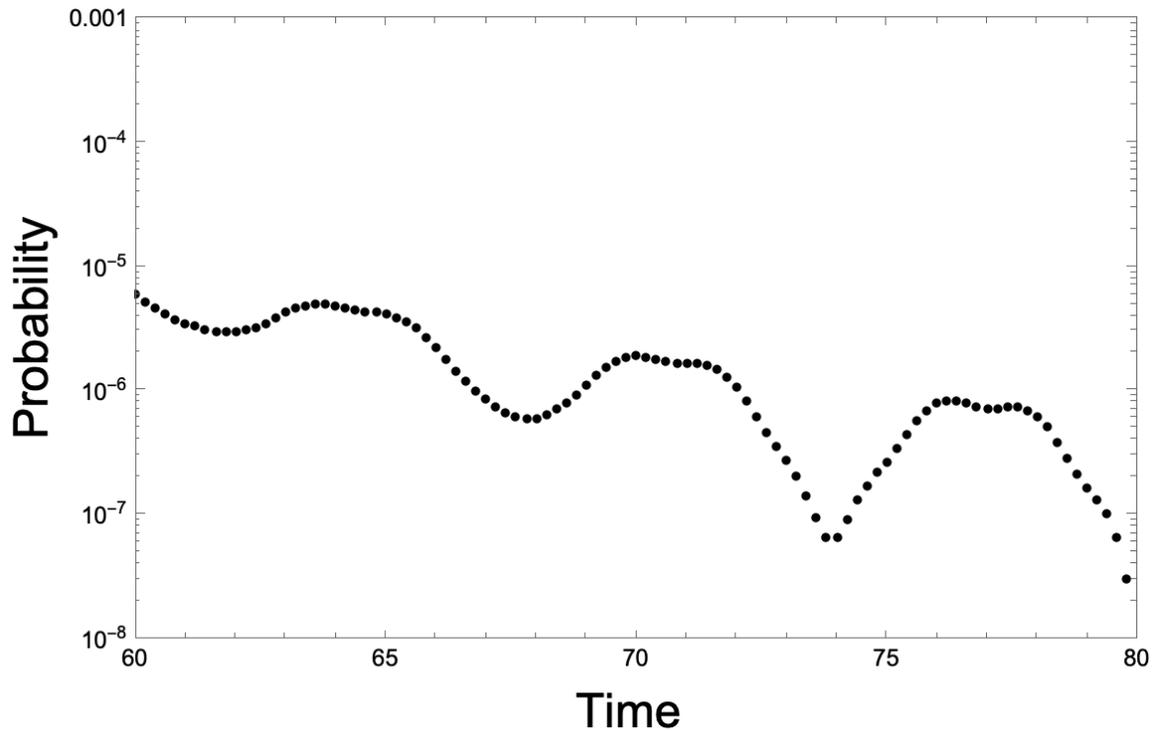

Figure 21. The survival probability versus time for the $\omega_{min} = -1$ and $\omega_{max} = 3$ continuum with $\alpha = 0.2, \bar{V} = 0.0$. The time step is 0.2 and note the vertical axis.

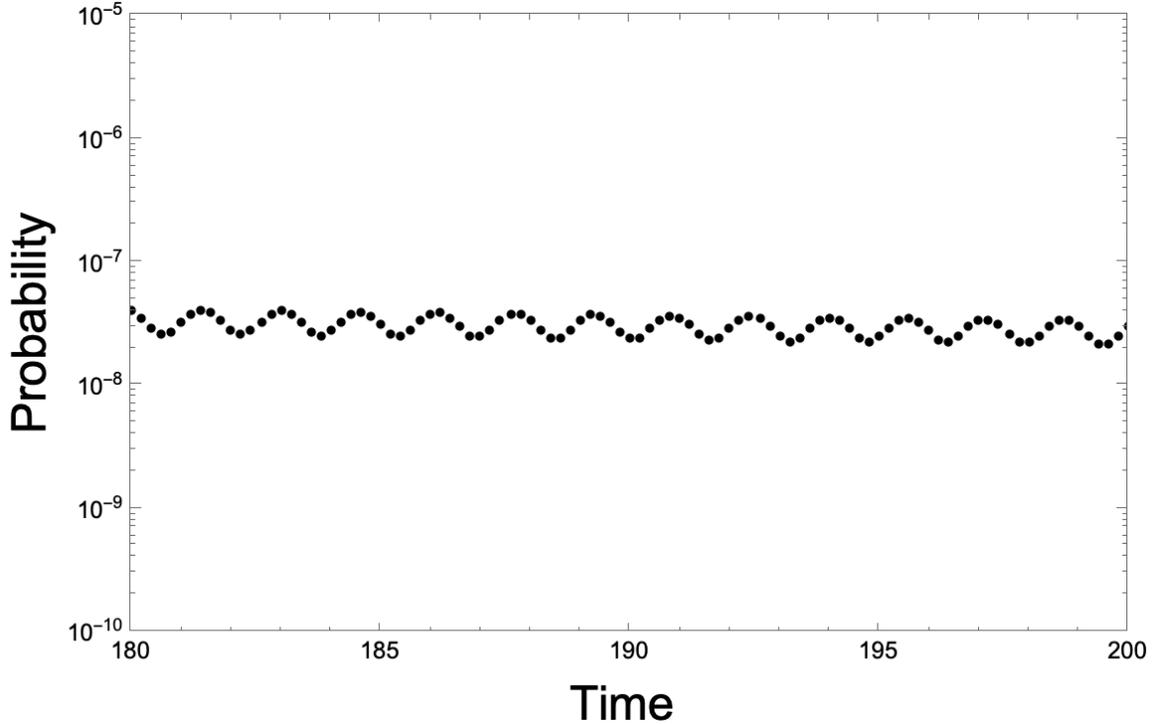

Figure 22. The survival probability versus time for the $\omega_{min} = -1$ and $\omega_{max} = 3$ continuum with $\alpha = 0.2, \bar{V} = 0.0$. The time step is 0.2 and note the vertical axis.

## 6. Conclusions and Discussion

This paper assumes an initially occupied state decays into a continuum of energy states and investigates whether the survival probability decreases in a monotonic fashion or not. The finite-state Bixon-Jortner model [5] yields a solution that has a sum over the states. This sum is turned into an integral over energy, Eq. (17), which is the complete solution of Schrödinger's Equation for an energy continuum that covers -∞ to +∞.

Three types of continua are explored using the above integral. The first has the energy run from -∞ to +∞ and the survival probability is an exponential in time for all times. Figure 3 demonstrates this for the first five decades of decay. Equations (8) and (9) then imply that no regrowth of any kind is allowed. This is consistent with the monotonic decrease with time of the survival probability when it is exponential in time.

The moments of the energy, $\langle \psi_s^0 | \hat{H}^n | \psi_s^0 \rangle$, are zero for $n$ odd and are infinite for $n$ even. Hence, the survival probability, $p_s(t)$, for small enough times $t$, should be linear in time according to Eq. (16), which comes from the first two terms of a power series expansion of the exponential for $p_s(t)$. Indeed, this is observed as is the constant slope with time.

The second continuum is truncated and it goes from $\omega_{min}$ to $+\infty$. The energy moments are infinite for both $n = 1$ and 2, so Eq. (16) is expected to be accurate and this is verified by Fig. 4. The initial decades of the decrease in $p_s(t)$ are not exponential in time, but the decrease becomes more of an exponential when $|\omega_{min}|$ increases. This trend in seen in Figs. 6 to 8. More striking, are the increases found in $p_s(t)$ in Figs. 6 and 7. The Paley-Wiener theorem discussed in Appendix A states that for a truncated continuum, the survival amplitude cannot be proportional to an exponential at large times. This is paired with Eq. (8) to say that regrowth occurs when the survival amplitude is not an exponential in time.

Hence, the regrowth or regeneration in $p_s(t)$ is expected. However, no cases are found where there is a $t$ such that $p_s(t > 0) = 1$ occurs or is even remotely approached. Now $p_s(t)$ is seen to drop several orders of magnitude before regrowth appears. When this is combined with the presence of the continuum, there is no significant build-up of the occupation probabilities of the energy states near the energy of the initial state. This is in contrast to what happens with a set of discrete energy states [11]. The observed regrowth is probed further for $\omega_{min} = -1$ starting with Fig. 9 for larger times. Oscillations in the survival probability are observed which decrease in amplitude and eventually show a power law decrease in Fig. 10. This dependence is in keeping with one of the possibilities discussed in Appendix A.

Now the present truncated continuum results are based on the use of a lower limit for the integral in Eq. (17). Figures 8 to 10 show this approximation captures the behavior of $p_s(t)$ found in the solution reported by Nakazato, Namiki, and Pascazio [20] and by Facchi [6] in his Ch. 4. This includes the exponential decay going over to a power law decay at large times.

The third continuum is truncated from below and from above, hence, it runs from $\omega_{min}$ to $\omega_{max}$. Now the energy moments are finite and Eq. (14)

should govern the small $t$ behavior. This is hinted at in Fig. 13 and confirmed in Fig. 14 where the slope of $p_s(t)$ is seen to go to zero as $t \to 0$. Larger times are portrayed in Figs. 15 and 17 for $\omega_{max} = |\omega_{min}| = 3$ and 1, respectively. The exponential decrease in $p_s(t)$ is evident especially in Fig. 15. A roughly exponential decay continues for even larger times for the $\omega_{max} = |\omega_{min}| = 3$ case with the oscillations in $p_s(t)$ growing in magnitude as the time increases. Figure 16 illustrates this. However, for $\omega_{max} = |\omega_{min}| = 1$, the oscillations start larger and continue with about the same peak magnitude as Figs. 18 to 20 make clear. When $\omega_{max} \neq |\omega_{min}|$, the survival probability has features of both of the $\omega_{max} = |\omega_{min}|$ cases illustrated.

In closing, there are a plethora of models that treat decay into a continuum of energy states. These range from Pietenpol's solution [15] for the full continuum and the solution represented by Eq. (17) and the present Appendix D, to the work of Nakazato, Namiki, and Pascazio [20] and Facchi [6] on the truncated continuum. It is worth trying to tame the solutions for the truncated continuum, since this continuum introduces a ground state. The approximate solutions treated in the present paper display a rich variety of behaviors for the survival probability versus time. It is of interest to see how many of these behaviors are found with more complete solutions. For example, the doubly-truncated continuum leads to a differential equation that is first-order in time when the procedure of Appendix D is applied. The right-hand side of this equation is an integral with an integrand that includes $A_s(t)$ and is under investigation.

Taken together, the answer to the title's question seems to be there is no complete recurrence but limited regrowth is observed.

# Appendix A: The large-time behavior of $A_s(t)$

Here I discuss two formal results for large times. The first shows the survival amplitude $A_s(t)$ goes to zero as $t$ goes to infinity, while the second proves $A_s(t)$ is not exponential as $t$ goes to infinity when the spectrum has a finite lower bound. The initial state is $|\psi_s^0\rangle$.

I follow Fonda, Ghirardi, and Rimini [4] and start with

$$A_s(t) = \langle \psi_s^0 | \hat{U}(t,0) | \psi_s^0 \rangle = \langle \psi_s^0 | \exp(-i\hat{H}t/\hbar) | \psi_s^0 \rangle, \quad (A1)$$

with the time-evolution operator for a time-independent Hamiltonian [7]. We let the eigenenergies and the eigenfunctions of $\hat{H}$ be

$$\hat{H} | \varphi(E) \rangle = E | \varphi(E) \rangle. \quad (A2)$$

Then

$$\int dE | \varphi(E) \rangle \langle \varphi(E) | = 1. \quad (A3)$$

We insert this into Eq. (A1), so

$$A_s(t) = \int dE \, \langle \psi_s^0 | \exp(-i\hat{H}t/\hbar) | \varphi(E) \rangle \langle \varphi(E) | \psi_s^0 \rangle = \quad (A4)$$

$$\int dE \, \exp(-iEt/\hbar) \langle \psi_s^0 | \varphi(E) \rangle \langle \varphi(E) | \psi_s^0 \rangle = \int dE \, \omega(E) \exp(-iEt/\hbar).$$

The last equality defines the spectrum density function $\omega(E)$ and note $\omega(E)$ depends on the initial state $\psi_s^0$.

Since the absolute value of the integral over the spectral density function

$$\int dE |\omega(E)| = \int dE |\langle \psi_s^0 | \varphi(E) \rangle|^2 = \int dE \, \langle \psi_s^0 | \varphi(E) \rangle \langle \varphi(E) | \psi_s^0 \rangle =$$

$$\langle \psi_s^0 | \psi_s^0 \rangle = 1, \quad (A5)$$

we may apply the Riemann-Lebesgue lemma [24] and find that

$$A_s(t) = \int dE \, \omega(E) \exp(-iEt/\hbar) \to 0 \text{ as } t \to \infty. \quad (A6)$$

This is our first result.

Now the limits of integration in the above may be infinite or finite. Most physical situations describe a system with a finite lower energy, say $E_{min} = \hbar \omega_{min}$, so our integrals over energy run from $E_{min}$ to $\infty$. A finite $E_{min}$ affects how $A_s(t)$ goes to zero as the time approaches infinity through a theorem due to Paley and Wiener [25]. This theorem says that if $\omega(E)$ vanishes for $E < E_{min}$, then its Fourier transform, defined in Eq. (A4), satisfies

$$\int_{-\infty}^{\infty} dt \, |ln|A_s(t)||/(1+t^2) < \infty. \tag{A7}$$

Equation (A7) rules out exponential decay as $t \to \infty$. Since, if the survival probability is exponential at large times, then

$$|A_s(t)| \to e^{-\gamma t}. \tag{A8}$$

This leads to the integral in Eq. (A7) going to

$$\gamma \int_{-\infty}^{\infty} dt \, |t|/(1+t^2) \to 2\gamma ln(1+t^2) \to \infty, \tag{A9}$$

when $t \to \infty$. This violates Eq. (A7) and rules out exponential decay at large times. This is a direct result of assuming the system has a ground state or a lower bound on its energy. A further consequence of the lower bound, and the non-exponential behavior of $A_s(t)$, is the non-zero second term in Eq. (8) of the main text. This means that even with a continuum of energy values some regeneration or regrowth will appear.

The Riemann-Lebesgue lemma has $A_s(t)$ approaching zero at large t, so $ln|A_s(t)|$ is negative at large t. This suggests two possible forms for $A_s(t)$ at large times. The first is

$$|A_s(t)| = e^{-ct^q}, \tag{A10}$$

with $c > 0$ and $q < 1$. If we define $q = m - 1$, and have $m < 2$, then the integral in Eq. (A7) becomes

$$2\int_0^{\infty} dt \, \{t^{m-1}/(1+t^2)\} = 2\pi/(sin(m\pi/2)), \tag{A11}$$

which is finite under the above restrictions. The second is a power law with

$$|A_s(t)| = 1/t^n, \tag{A12}$$

with $n > 0$. The integral in Eq. (A7) is now [16]

$$2 \int_0^\infty dt |lnt|/(1 + t^2) = 4(0.91596 \ldots),  \qquad (A13)$$

and the number in parenthesis is Catalan's constant. So, the result is again finite.

I have taken the large-time behavior for all times in order to do the integrals and I do so to provide a guide as to how $A_s(t)$ behaves at large times.

## Appendix B: The Bixon-Jortner Model

The model [5] is developed for a finite set of states for ease. We start with a state denoted by the subscript $s$ with a population of 1 and with all the other states unoccupied. State $s$ decays into a set of states that are labelled by $k = \{-m, -m+1, \ldots, -1, 0, 1, \ldots, m-1, m\}$. The matrix element for a transition from state $s$ to any state $k$ is

$$\bar{V} = \langle k|\hat{V}|s\rangle/\hbar = \langle s|\hat{V}|k\rangle/\hbar, \qquad (B1)$$

and the matrix element is real and a constant that is the same for all $k$. The hat indicates an operator. There are no transitions between any of the $k$-states. Next, we need a Hamiltonian. Let

$$\hat{H} = \widehat{H_0} + \hat{V}, \qquad (B2)$$

and

$$\widehat{H_0}|i\rangle = E_i|i\rangle. \qquad (B3)$$

Here $i$ is $s$ or a $k$. These states are assumed to be orthonormal. This summarizes the Bixon-Jortner model.

The model is deceptively simple-looking. Solutions require numerical methods and this appendix uses a matrix representation of the Hamiltonian. The sparse structure of this matrix allows the eigenvectors to be found in a simple form. The eigenvalues arise from an equation involving the cotangent. This method leads to a sum over $s$ and all the $k$-states as follows.

Let

$$E_i = \hbar\omega_i, \tag{B4}$$

denote the eigenvalues of $\widehat{H}_0$, while the eigenvalues of $\widehat{H}$ are written

$$E_j = \hbar\omega'_j. \tag{B5}$$

The time-dependent ket is now expressed in terms of the eigenfunctions of the complete Hamiltonian $\widehat{H}$, the $\phi_j$ with

$$\phi_j = a_j \psi_s^0 + \sum_k b_k^j \psi_k^0, \tag{B6}$$

and, yes, the superscript 0 indicates these functions come from $\widehat{H}_0$. The interest is in the survival amplitude of the initial state $|\psi_s^0\rangle$, which is written with the aid of the time-evolution operator, $\widehat{U}(t, t_0 = 0)$, as

$$A_s(t) = \langle \psi_s^0 | \widehat{U}(t,0) | \psi_s^0 \rangle = \langle \psi_s^0 | exp(-i\widehat{H}t/\hbar) | \psi_s^0 \rangle. \tag{B7}$$

The equality follows since $\widehat{H}$ is independent of time [7]. This expression is simplified by inserting a complete set of states of $\widehat{H}$ between the operator and the ket. The set index $j$ includes $s$ and the $2m+1$ states for the sum over $k$, so

$$A_s(t) = \sum_j \langle \psi_s^0 | exp(-i\widehat{H}t/\hbar) | j \rangle \langle j | \psi_s^0 \rangle = \sum_j \langle \psi_s^0 | j \rangle exp(-i\omega'_j t) \langle j | \psi_s^0 \rangle. \tag{B8}$$

Each state $j$ is based on Eq. (B6), hence, using the orthonormality of the kets in Eq. (B3), which are written in Eq. (B6) with the superscript 0, Eq. (B8) reduces to

$$A_s(t) = \sum_j a_j^* a_j exp(-i\omega'_j t). \tag{B9}$$

The initial state occupation probability, the survival probability $p_s(t)$, is found by multiplying $A_s(t)$ by its complex conjugate. Thus, we need to find the $a_j$. These follow from the matrix representation of the Hamiltonian $\widehat{H}$.

For the $j$ th eigenvalue $\omega_j'$ of $\hat{H}$, $a_j$ and the set of $\{b_k^j\}$ form its eigenvector and are found from

$$\begin{pmatrix} \omega_s & \bar{V} & \bar{V} & . & \bar{V} & . \\ \bar{V} & \omega_1 & 0 & . & 0 & . \\ \bar{V} & 0 & \omega_2 & . & 0 & . \\ . & . & . & . & . & . \\ \bar{V} & 0 & 0 & . & \omega_k & . \\ . & . & . & . & . & . \end{pmatrix} \begin{pmatrix} a_j \\ \vdots \\ \vdots \\ b_k^j \\ \vdots \end{pmatrix} = \omega_j' \begin{pmatrix} a_j \\ \vdots \\ \vdots \\ b_k^j \\ \vdots \end{pmatrix} . \tag{B10}$$

Equation (B10) produces the equations

$$\omega_s a_j + \bar{V} \sum_k b_k^j = \omega_j' a_j , \tag{B11}$$

and

$$\bar{V} a_j + \omega_k b_k^j = \omega_j' b_k^j . \tag{B12}$$

The last equation gives

$$b_k^j = - \bar{V} a_j / (\omega_k - \omega_j') , \tag{B13}$$

and this is substituted into Eq. (B11), so that

$$(\omega_s - \omega_j') a_j - a_j \overline{VV} \sum_k \frac{1}{(\omega_k - \omega_j')} = 0 . \tag{B14}$$

The factor $a_j$ is dropped and the eigenvalue $\omega_j'$ is found from

$$(\omega_s - \omega_j') - \bar{V}^2 \sum_k \frac{1}{(\omega_k - \omega_j')} = 0 . \tag{B15}$$

The next step is to relate the sum in Eq. (B15) to the cotangent. This starts with the assumption that the eigenvalues of $\hat{H}_0$ are written as

$$\omega_k = \omega_s + k\varepsilon , \tag{B16}$$

with $\varepsilon$ a parameter giving the unperturbed energy level separation. Now defining

$$x_j = (\omega'_j - \omega_s)/\varepsilon , \tag{B17}$$

causes the denominator in the sum in Eq. (B15) to become

$$(\omega_k - \omega'_j) = -\frac{\varepsilon(\omega'_j - \omega_s - k\varepsilon)}{\varepsilon} = -\varepsilon(x_j - k) , \tag{B18}$$

and Eq. (B15) is rewritten as

$$(\omega_s - \omega'_j) + (\bar{V}^2/\varepsilon) \sum_k \frac{1}{(x_j - k)} = 0 . \tag{B19}$$

The sum is known if it runs over $k = -\infty$ to $+\infty$ and in that case

$$\sum_k \frac{1}{(x_j - k)} = \pi \cot \pi x_j . \tag{B20}$$

This surprising result is explained in Appendix C. Its use here amounts to an additional source of error when a sum is used. The combination of Eqs. (B19) and (B20) makes the equation for the eigenvalues of $\hat{H}$

$$(\omega_s - \omega'_j) + (\bar{V}^2/\varepsilon)\pi \cot \pi x_j = 0 , \tag{B21}$$

with $\omega'_j$ in $x_j$.

The eigenfunctions of Eq. (B6) are normalized by

$$(a_j)^2 + \sum_k (b_k^j)^2 = 1 , \tag{B22}$$

and this is simplified by the use of Eqs. (B13) and (B17) to

$$(a_j)^2 + \left(\frac{\bar{V}}{\varepsilon}\right)^2 (a_j)^2 \sum_k \frac{1}{(x_j - k)^2} = 1 . \tag{B23}$$

The index $k$ is allowed to run from $-\infty$ to $+\infty$ in order to take advantage of Eq. (B20). Then the sum is equal to the derivative of the sum with linear terms in the denominator and

$$\pi \frac{d}{dx_j} cot\pi x_j = -\frac{\pi^2}{(sin\pi x_j)^2} = -\pi^2\left(1 + (cot\pi x_j)^2\right). \quad (B24)$$

This result is combined with Eq. (B21) to yield

$$(a_j)^2 = \bar{V}^2/\left(\bar{V}^2 + (\omega_s - \omega_j')^2 + (\pi\bar{V}^2/\varepsilon)^2\right). \quad (B25)$$

If we pass from a finite set of states to a continuum of energy states, then Eq. (17) results.

# APPENDIX C: Infinite sum and the cotangent

The infinite sum is turned into a cotangent as follows:

$$\Sigma_k \frac{1}{(z-k)} = \frac{1}{z} + \Sigma_{k=1} \frac{1}{(z-k)} + \Sigma_{k=-1} \frac{1}{(z-k)}, \quad (C1)$$

and the sums run to $+\infty$ and $-\infty$, respectively. Let $k \to -k$ in the second sum and now both sums run over 1 to $+\infty$. The terms for a fixed $k$ are

$$\frac{1}{(z-k)} + \frac{1}{(z+k)} = \frac{2z}{(z^2-k^2)}. \quad (C2)$$

Equation (C1) becomes

$$\Sigma_k \frac{1}{(z-k)} = \frac{1}{z} + \Sigma_{k=1} \frac{2z}{(z^2-k^2)} = \pi cot\pi z, \quad (C3)$$

which is known as the Mittag-Leffler Expansion [26].

## Appendix D: The Solution for the Full Continuum

I follow Section 3.4 of [6]. Section 2 explains which transition matrix elements are non-zero. These are embodied in the following Hamiltonian

$$\hat{H} = \omega_s |s\rangle\langle s| + \int d\omega \omega |\omega\rangle\langle\omega| + \lambda \int d\omega (|s\rangle\langle\omega| + |\omega\rangle\langle s|) , \quad (D1)$$

with the initial state denoted by $|s\rangle$. The ket at time $t$ is

$$|\psi(t)\rangle = a(t)|s\rangle + \int d\omega b(\omega,t) |\omega\rangle , \quad (D2)$$

and we need to find $a(t)$ and $b(\omega,t)$. We do this through Schrodinger's Equation, which with $\hbar = 1$ is

$$i\frac{\partial}{\partial t}|\psi(t)\rangle = i\dot{a}(t)|s\rangle + i\int d\omega \dot{b}(\omega,t)|\omega\rangle = \hat{H}|\psi(t)\rangle , \quad (D3)$$

where the dot indicates the time derivative.

Let us work on the right-hand side of Eq. (D3). When the Hamiltonian acts on the first term on the right-hand side of Eq. (D2), we find the non-zero terms are

$$\omega_s a(t)|s\rangle + \lambda a(t) \int d\omega |\omega\rangle . \quad (D4)$$

Similarly, the Hamiltonian and the second term contribute

$$\int d\omega \omega b(\omega,t) |\omega\rangle + \int d\omega b(\omega,t) |a\rangle. \quad (D5)$$

We now have all the non-zero terms in Eq. (D3) and we multiply from the left by $\langle s|$ and discover

$$i\dot{a}(t) = \omega_s a(t) + \lambda \int d\omega b(\omega,t) . \quad (D6)$$

Next, we multiply Eq. (D3) by $\langle\omega'|$ and find

$$i\dot{b}(\omega,t) = \lambda a(t) + \omega b(\omega,t) . \quad (D7)$$

So, we have two coupled equations and we start by realizing Eq. (D7) is a first-order, linear, ordinary differential equation for $b(\omega, t)$ as a function of $t$. Hence, the standard solution [27] is

$$ib(\omega, t) = \lambda \int_0^t dt' a(t') \exp[-i\omega(t - t')] \,. \tag{D8}$$

We place this solution into Eq. (D6) and integrate $\omega$ over $-\infty$ to $+\infty$, this allows us to get a delta function in time and Eq. (D6) becomes

$$\dot{a}(t) = (-i\omega_s - \lambda^2 \pi) a(t) \,, \tag{D9}$$

and this produces an exponential in time

$$a(t) = \exp(-i\omega_s t) \exp(-\lambda^2 \pi t) \,. \tag{D10}$$

We have arrived at Eq. (18)!

## Appendix E: Integrals for the Doubly-Truncated Continuum

Here I treat the case with $\omega_{max} = |\omega_{min}|$, so the sine integral in Eq. (17) is zero. Now for the cosine integral of Eq. (17). First, the denominator is split into partial fractions via

$$\frac{1}{(\omega^2 + \alpha^2)} = (i/2\alpha)\{(1/(\omega + i\alpha)) - (1/(\omega - i\alpha))\} \,. \tag{E1}$$

Next, with $\alpha = 0.2$, Mathematica [16] provides

$$\int d\omega \{\cos(\omega t)/(\omega + i\alpha)\} = \cosh(\alpha t)\, CosIntegral(\alpha t i + \omega t) + i\sinh(\alpha t) SinIntegral(\alpha t i + \omega t) \,, \tag{E2}$$

and

$$\int d\omega \{\cos(\omega t)/(\omega - i\alpha)\} = \cosh(\alpha t)\, CosIntegral(\alpha t i - \omega t) + i\sinh(\alpha t) SinIntegral(\alpha t i - \omega t) \,. \tag{E3}$$

These last two equations are combined with Eq. (E1) and integrated over $\omega$

as a function of time. Then their sum is multiplied by its complex conjugate to get a probability. The following figure shows the unnormalized results for $\omega_{max} = 1$ and $\alpha = 0.2$. The oscillations are evident.

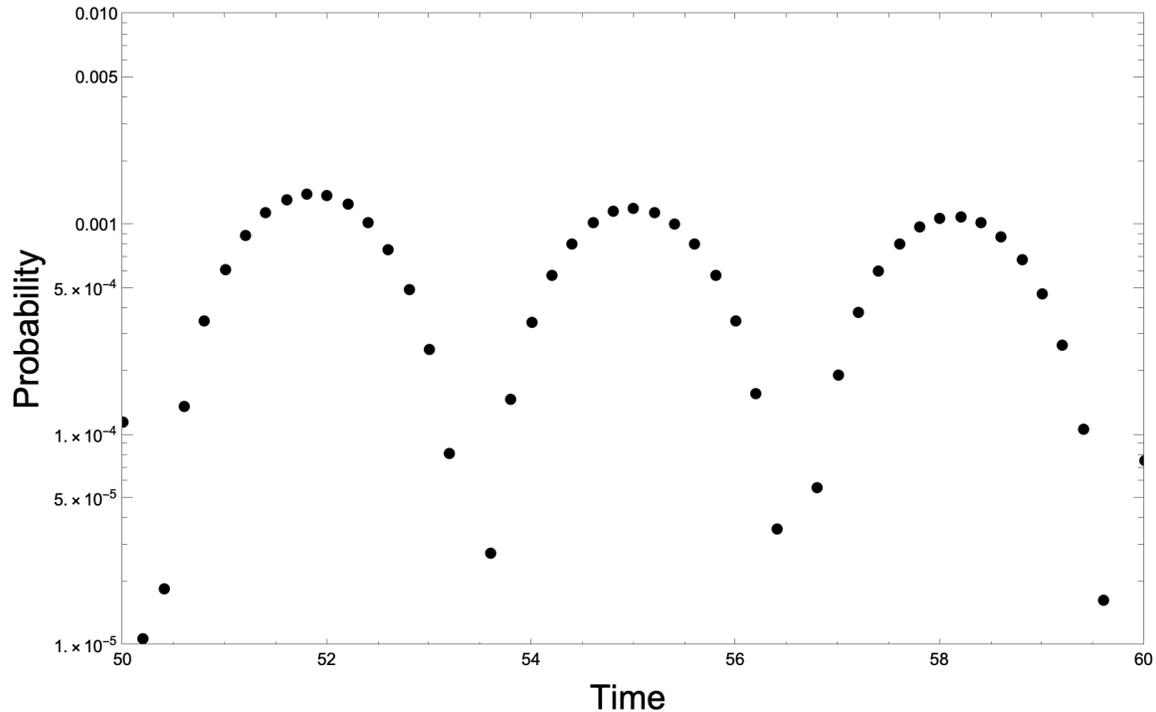

Figure E1. The oscillating contribution for the doubly-truncated continuum with $\omega_{max} = 1 = |\omega_{min}|$ and $\alpha = 0.2$. The time step is 0.2.

## Acknowledgments

I thank Harvey S. Picker for discussions and communications about the role of continua in decays.

## References

[1] P. Bocchieri and A. Loinger, "Quantum recurrence theorem," Phys. Rev. **107**, 337-338 (1957).

[2] L. S. Schulman, "Note on the quantum recurrence theorem," Phys. Rev. A**18**, 2379-2380 (1978).

[3] L. S. Schulman, *Time's arrow and quantum measurement* (Cambridge University Press, Cambridge, 1997), Ch. 2.

[4] L. Fonda, G. C. Ghirardi, and A. Rimini, "Decay theory of unstable